\documentclass[%
 aip,
 jcp,
 sd,%
 amsmath,amssymb,
 reprint,%
]{revtex4-1}

\usepackage{graphicx}
\usepackage{amsfonts}
\usepackage{amsmath}
\usepackage{mathtools}
\usepackage{float}
\usepackage{epsfig}
\usepackage{color}
\usepackage{gensymb}
\usepackage{multirow}
\usepackage[mathscr]{euscript}

\newcommand{\e}[1]{\ensuremath{\times 10^{#1}}}
\newcommand{\ee}[1]{\ensuremath{10^{#1}}}

\newcommand{\Skip}[1]{}

\begin{document}

\title{Transport, Geometrical and Topological Properties of Stealthy Disordered Hyperuniform Two-Phase Systems}

\author{G. Zhang}


\affiliation{\emph{Department of Chemistry}, \emph{Princeton University},
Princeton NJ 08544}

\author{F. H. Stillinger}


\affiliation{\emph{Department of Chemistry}, \emph{Princeton University},
Princeton NJ 08544}

\author{S. Torquato}

\email{torquato@electron.princeton.edu}

\affiliation{\emph{Department of Chemistry, Department of Physics,
Princeton Institute for the Science and Technology of
Materials, and Program in Applied and Computational Mathematics}, \emph{Princeton University},
Princeton NJ 08544}


\begin{abstract}
Disordered hyperuniform many-particle systems have attracted considerable recent attention, since they behave like crystals in the manner in which they suppress large-scale density fluctuations, and yet also resemble statistically isotropic liquids and glasses with no Bragg peaks.
One important class of such systems is the classical ground states of ``stealthy potentials.'' The degree of order of such ground states depends on a tuning parameter $\chi$. Previous studies have shown that these ground-state point configurations can be counterintuitively disordered, infinitely degenerate, and endowed with novel physical properties (e.g., negative thermal expansion behavior).
In this paper, we focus on the disordered regime ($0<\chi<1/2$) in which there is no long-range order, and control the degree of short-range order. 
We map these stealthy disordered hyperuniform point configurations to two-phase media by circumscribing each point with a possibly overlapping sphere of a common radius $a$: the ``particle'' and ``void'' phases are taken to be the space interior and exterior to the spheres, respectively.
The hyperuniformity of such two-phase media depend on the sphere sizes:
While it was previously analytically proven that the resulting two-phase media maintain hyperuniformity if spheres do not overlap, here we show numerically that they lose hyperuniformity whenever the spheres overlap.
We study certain transport properties of these systems, including the effective diffusion coefficient of point particles diffusing in the void phase
as well as static and time-dependent characteristics associated with diffusion-controlled reactions. 
Besides these effective transport properties, we also investigate several related structural properties, including pore-size functions, quantizer error, an order metric, and percolation thresholds. We show that these transport, geometrical and topological properties of our two-phase media derived from decorated stealthy ground states are distinctly different from those of equilibrium hard-sphere systems and spatially uncorrelated overlapping spheres. 
As the extent of short-range order increases, stealthy disordered two-phase media can attain nearly maximal effective diffusion coefficients over a broad range of volume fractions while also maintaining isotropy, and therefore may have practical applications in situations where ease of transport is desirable. We also show that the percolation threshold and the order metric are positively correlated with each other, while both of them are negatively correlated with the quantizer error. In the highly disordered regime ($\chi \rightarrow 0$), stealthy point-particle configurations are weakly-perturbed ideal gases. Nevertheless, reactants of diffusion-controlled reactions decay much faster in our two-phase media than in equilibrium hard-sphere systems of similar degrees of order, and hence indicate that the formation of large holes is strongly suppressed in the former systems.

\end{abstract}

\maketitle
 
\section{Introduction}

A hyperuniform many-particle system is one in which the structure factor approaches zero in the infinite-wavelength limit.\cite{torquato2003local} In such systems, density fluctuations (measured by the variance of number of particles inside a spherical window) are anomalously suppressed at very large lengths scales, a ``hidden'' order that imposes strong global structural constraints.\cite{torquato2003local, torquato2016hyperuniformity} All structurally perfect crystals and quasicrystals are hyperuniform,\cite{torquato2003local, zachary2009hyperuniformity} but typical disordered many-particle systems, including gases, liquids, and  glasses, are not. 
Disordered hyperuniform many-particle systems are exotic states of amorphous matter that have attracted considerable recent attention.\cite{torquato2003local, zachary2009hyperuniformity, donev2005unexpected, zachary2011hyperuniform, jiao2011maximally, atkinson2016critical, kurita2011incompressibility, dreyfus2015diagnosing, lesanovsky2014out, hexner2015hyperuniformity, jack2015hyperuniformity, de2015toward, degl2015thz, xie2013hyperuniformity, muller2014silicon, florescu2009designer, torquato2015ensemble, uche2004constraints, zhang2015ground, zhang2015ground2, batten2009novel, novikov2014revealing, torquato2016hyperuniformity, xu2016influence}
Materials that are simultaneously disordered and hyperuniform can be regarded to be exotic states of matter that lie between a crystal and a liquid; they behave more like crystals in the manner in which they
suppress large-scale density fluctuations, and yet they also
resemble typical statistically isotropic liquids and glasses with no Bragg peaks.\cite{torquato2015ensemble}

An important class of disordered hyperuniform many-particle systems is comprised of the classical ground states of ``stealthy potentials,''\cite{uche2004constraints, torquato2015ensemble, zhang2015ground, zhang2015ground2} which are bounded, long-range, pairwise additive potentials designed in Fourier space. These classical ground states are of particular fundamental interest because they can be degenerate and noncrystalline. A nonnegative parameter inversely proportional to the number density, $\chi$, controls the degree of order of such ground states. For $\chi<0.5$, the ground states are overwhelmingly highly degenerate and disordered. As $\chi$ increases above 0.5, long-range translational and rotational order begins to emerge and eventually the system crystallizes. We have previously studied these disordered ground states, and computed their pair correlation functions,\cite{uche2004constraints, uche2006collective, batten2008classical, zhang2015ground, torquato2015ensemble} structure factors,\cite{uche2004constraints, uche2006collective, batten2008classical, zhang2015ground, torquato2015ensemble} Voronoi cell volume distribution,\cite{uche2004constraints, zhang2015ground} and particle-exclusion probabilities.\cite{torquato2015ensemble}

Some initial studies have demonstrated that stealthy hyperuniform systems are endowed with novel thermodynamic and physical properties. For example, their low-temperature excited states are characterized by negative thermal expansion behavior.\cite{batten2009novel}
It has also been shown that dielectric networks derived from stealthy disordered hyperuniform point configurations possess complete photonic band gaps comparable in size to those of a photonic crystal, while at the same time
maintain statistical isotropy, enabling waveguide geometries not possible with photonic crystals as well as high-density disordered transparent materials.\cite{florescu2009designer, florescu2013optical, man2013isotropic, leseur2016high} 
However, the determination of physical/chemical properties of stealthy disordered hyperuniform materials is generally an unexplored area of research.

In this paper, we investigate steady-state and time-dependent diffusion properties of certain decorations of stealthy disordered hyperuniform ground-state point configurations in two and three dimensions.
In particular, we derive two-phase heterogeneous media from point configurations by decorating the point configurations with spheres (circles); specifically, all points are circumscribed by spheres of radius $a$ that generally may overlap with one another. By varying the radius, the fraction of space occupied by the spheres will vary. We study the effective transport properties of these disordered two-phase systems, including the effective diffusion coefficient,\cite{torquato2013random} and static and time-dependent characteristics of diffusion-controlled reactions at the interfaces between the two continuous phases, as well as the trapping rate (or its inverse, the mean survival time) as well as the principal (largest) relaxation time.\cite{reck1965diffusion, torquato1991diffusion} Quantifying the effective diffusion coefficient is of importance not only because it has direct applications (e.g., diffusion of fuel and oxygen in a fuel cell \cite{zamel2010experimental}, diffusion tensor magnetic resonance imaging, \cite{sen1989effective, tuch2001conductivity} regulation and metabolism of normal organs,  \cite{comper1996extracellular, gevertz2008novel} and drug release from porous matrices \cite{lemaire2003structural}), but also because its determination translates immediately into equivalent results for the effective thermal and electric conductivity, the effective dielectric constant, and the effective magnetic permeability for reasons of mathematical analogy,\cite{torquato2013random} and is therefore related to a host of applications. Diffusion-controlled reactions arise in widely different processes, such as heterogeneous catalysis,\cite{baiker1999supercritical} gas sensor operation,\cite{sakai2001theory} cell metabolism,\cite{rohde2007diffusion} crystal growth,\cite{watson2009non} and nuclear magnetic resonance (NMR).\cite{banavar1987magnetic, mitra1992effects, straley1987magnetic}

These transport properties are related to several statistical geometrical and topological characteristics, which we therefore also study. These include the pore-size functions (the distribution of the distance from a randomly chosen location in the void phase to the closest phase boundary),\cite{torquato2010reformulation} the quantizer error (a moment of the pore-size function, which is related to the principal relaxation time),\cite{torquato1991diffusion, torquato2010reformulation} the order metric $\tau$ (a measure of the translational order of point configurations),\cite{torquato2015ensemble} and the percolation threshold or the critical radius (the radius of the spheres at which a specific phase becomes connected) of each phase.\cite{rintoul1997precise, lorenz2001precise, quintanilla2000efficient, quintanilla2007asymmetry}

We compare the aforementioned physical and geometrical properties of our two-phase system derived from decorated stealthy ground states, as a function of the tuning parameter $\chi$, with those of two other two-phase media: (1) equilibrium disordered (fluid) hard-sphere systems and (2) decorated Poisson point processes (ideal-gas configurations). The former has short-range order that is tunable by its volume fraction but no long-range order. The latter has neither short-range order nor long-range order. Through comparison, we find that some of these quantities are dramatically affected by the degree of long-range order, while other quantities are much more sensitive to the degree of short-range order. Because many of these quantities depend on the density, we re-scale all systems to unit number density to ensure a fair comparison.

Among our major findings, we show that these transport, geometrical and topological properties of our two-phase media are generally distinctly different from those of equilibrium hard-sphere systems and spatially uncorrelated overlapping spheres. 
At high $\chi$ values, the stealthy disordered two-phase media can attain nearly maximal effective diffusion coefficient, while also maintaining isotropy. This novel property could have practical implications, e.g., optimal and isotropic drug release from designed  nanoparticles. 
Stealthy ground states tend to ideal gases configurationally in the $\chi \to 0$ limit. \cite{torquato2015ensemble}
Nevertheless, we find that even in the low-$\chi$ regime, our two-phase media have much lower principal relaxation time than that of equilibrium hard-sphere systems of similar degrees of order, indicating that the formation of large holes in the stealthy systems is strongly suppressed.
Lastly, we also find that the aforementioned geometrical and topological quantities are strongly correlated with each other.

The rest of the paper is organized as follows: In Sec.~\ref{definition}, we give precise definitions of the stealthy potential and the aforementioned
transport, geometrical and topological quantities. In Sec.~\ref{simulation}, we present our numerical method to calculate them. We present our results in Sec.~\ref{result} and conclusions in Sec.~\ref{conclusion}.

\section{Mathematical definitions and background}
\label{definition}

\subsection{Preliminaries}

This paper studies properties of point-particle systems as well as two-phase  heterogeneous media derived from certain decorations of these point configurations.
A point-particle system consists of $N$ point particles with a certain probability density function $P(\mathbf r^N)$, where $\mathbf r^N \equiv \mathbf r_1$, $\mathbf r_2$, ..., $\mathbf r_N$ is the particle positions, in a simulation box of volume $v_F$ under periodic boundary conditions in $d$-dimensional Euclidean space $\mathbb{R}^d$, where $d$ is 2 or 3. The number density is defined as $\rho=N/v_F$. The ``Poisson point process'' (also called ``ideal gas'') is produced by the probability density function $P(\mathbf r^N)=v_F^{-N}$ that does not depend on particle positions $\mathbf r^N$. The equilibrium hard-sphere point process of radius $a$ is another point process with $P(\mathbf r^N)$ equal to a positive constant if the distance between every pair of points is larger than $2a$ and zero otherwise.

A realization of a two-phase medium can be mathematically described as a partition of a domain of space $\mathscr V \in \mathbb R^d$ with volume $V$ into two separate regions, $\mathscr V_1$ and $\mathscr V_2$.  It is characterized by an indicator function, $\mathcal I(\mathbf x)$, where $\mathbf x$ is any position in the two-phase medium. The indicator function $\mathcal I(\mathbf x)$ is one if $\mathbf x \in \mathscr V_1$ and zero if $\mathbf x \in \mathscr V_2$. The volume fraction of phase 1 is given by $\phi_1=<\mathcal I(\mathbf x)>$, where $<\cdots>$ denotes an ensemble average. That of the other phase is given by $\phi_2=1-\phi_1$. Let $\partial \mathscr V$ be the interface between $\mathscr V_1$ and $\mathscr V_2$, the specific surface, i.e., the total area of $\partial \mathscr V$ divided by $V$, is given by:
\begin{equation}
s=<|\nabla \mathcal I(\mathbf x)|>.
\end{equation} 

The two-phase media that we consider here are derived from point configurations by decorating the point configurations with spheres (circles); specifically, each point is circumscribed by a sphere of radius $a$ that generally may overlap with one another. Therefore, it is composed of a void region (phase 1) and a particle region (phase 2). 
When such a mapping is applied to a Poisson point process, the decorated system is also called ``fully penetrable spheres'' \cite{quintanilla2007asymmetry} or ``spatially uncorrelated spheres.'' \cite{lorenz2001precise} 


\subsection{Stealthy potentials and their entropically favored ground states}
\label{subsection_StealthyPotential}
Consider point processes that are obtained from the canonical ensemble probability distribution function defined by
\begin{equation}
P(\mathbf r^N)=\exp[-\beta \Phi(\mathbf r^N)]/Z,
\label{canonical}
\end{equation} 
where $\Phi(\mathbf r^N)$ is an interaction potential, $\beta$ is the inverse temperature, and $Z=\int \exp[-\beta \Phi(\mathbf r^N)] d \mathbf r^N$ is the partition function. Of particular interest in this paper is the ``stealthy'' interaction potential:
\begin{equation}
\begin{split}
\Phi({\bf r}^N) &=\frac{1}{2v_F} \sum_{0< \mathbf k < K} |{\tilde n}({\bf k})|^2 +\Phi_0 \\
&= \sum_{i<j} \frac{1}{v_F} \sum_{0< \mathbf k < K} \exp( i \mathbf k \cdot \mathbf r_{ij}),
\end{split}
\label{pot_f}
\end{equation}
where the sum is over all reciprocal lattice vector $\mathbf k$'s of the simulation box such that $0<|\mathbf k|\le K$, ${\tilde n}({\bf k})= \sum_{j=1}^{N} \exp(-i{\bf k \cdot r}_j)$, 
\begin{equation}
\Phi_0=[N(N-1) - \sum_{0< \mathbf k < K} N ]/2v_F
\label{temp1}
\end{equation}
is a constant independent of the particle positions $\mathbf r^N$, and the second equal sign in Eq.~(\ref{pot_f}) can be proved by Parseval's theorem. Such potential is interesting not only because it is a pairwise additive potential [as the right side of Eq.~(\ref{pot_f}) shows], but also because it allows one to directly tune the structure factor 
\begin{equation}
S(\mathbf k)=|{\tilde n}({\mathbf k})|^2/N.
\label{structurefactor}
\end{equation}
The ground state (i.e., $\beta \to + \infty$ or zero-temperature state) of this potential is obtained by constraining $S(\mathbf k)=0$ for all $0<|\mathbf k|\le K$.\cite{uche2004constraints, torquato2015ensemble}

Let $M$ be half the number of $\mathbf k$ points in the summation of Eq.~\eqref{pot_f} \footnote{Since $|{\tilde n}({\bf k})|^2=|{\tilde n}({-\bf k})|^2$, $M$ is the number of independent constraints.
}; the parameter
\begin{equation}
\chi=\frac{M}{d(N-1)}
\label{chi}
\end{equation}
determines the degree to which the ground states are constrained and therefore the degeneracy and disorder of the ground states.\cite{uche2004constraints} For $\chi<0.5$, the ground states are typically disordered and uncountably infinitely degenerate.\cite{torquato2015ensemble, zhang2015ground} Therefore, there are multiple ways to assign different weights (i.e., probabilities) to different sets of ground states. One way of particular interest is the zero-temperature ($\beta \to + \infty$) limit of Eq.~(\ref{canonical}). Ground states drawn from such distribution are called ``entropically favored ground states''.\cite{torquato2015ensemble, zhang2015ground} It is interesting to note that in the $\chi \to 0$ and $a \to 0$ limit, both entropically favored ground states of stealthy potentials and equilibrium hard-sphere point processes tend to Poisson point process geometrically. In the rest of the paper this fact will be frequently used to test our simulation results since many properties of the Poisson point process have been studied previously.

\subsection{Transport properties}

This paper studies the following steady-state and time-dependent diffusion properties in phase 1 (the void phase) of decorated entropically favored ground states of stealthy potentials, and compare them with that of decorated Poisson point process and equilibrium disordered (fluid) hard-sphere system at unit number density.

\subsubsection{Effective diffusion coefficient} 
Consider the steady-state diffusion problem of some species with concentration field $c(\mathbf x)$ in a two-phase medium in which phase 1 is the space in which diffusion occurs and phase 2 are ``obstacles'' that the diffusing species cannot enter. In phase 1, the flux of the species, $\mathbf J(\mathbf x)$, is predicted by Fick's first law:
\begin{equation}
\mathbf J(\mathbf x) = D \nabla c(\mathbf x)\mbox{, $\mathbf x \in \mathscr V_1$}
\label{eq_diffusion}
\end{equation}
where $D$ is a diffusion coefficient which we set to unity for simplicity. However, Eq.~(\ref{eq_diffusion}) is valid only in phase 1 and has to be paired with the following Neumann boundary condition:
\begin{equation}
\mathbf n \cdot \mathbf J =0\mbox{, on $\partial \mathscr V$,}
\end{equation}
where $\mathbf n$ is the normal vector of the surface. We see that the inclusion of such obstacles adds a complicated boundary condition and makes the overall diffusion problem difficult. Nevertheless, on a length scale much larger than the characteristic length of the obstacles, the system can be homogenized \cite{torquato2013random} and characterized by an ``effective'' diffusion coefficient, $D_{e}$, defined by the average Fick's first law:
\begin{equation}
<\mathbf J(\mathbf x)> = D_{e} <\nabla c(\mathbf x)>\mbox{, for any $\mathbf x$}
\end{equation}
where angular brackets denote ensemble averages.

The effective diffusion coefficient of an isotropic two-phase medium must satisfy
the Hashin-Shtrikman (HS) upper bound.\cite{hashin1963variational} For our case where phase 1 has unit diffusion coefficient and phase 2 cannot be entered, this bound in $d$ dimensions is given by:
\begin{equation}
D_e \le \frac{d-1}{d-1+\phi_2}.
\label{HSBound}
\end{equation}
This bound is optimal because it is realizable by certain model microstructures, including the ``coated-sphere model'' described in Ref.~\onlinecite{hashin1962elastic}, and is therefore the best possible bound for isotropic systems given volume-fraction information only.

\subsubsection{Diffusion-controlled reactions} Consider the problem of diffusion and reaction among absorbing ``traps'' in the random medium. Let phase 1 be the region in which diffusion occurs and phase 2 be the trap region, the diffusion process in phase 1 is governed by the same Fick's first law but with time dependency: 
\begin{equation}
\mathbf J(\mathbf x, t) = D \nabla c(\mathbf x, t), \mbox{ in $\mathscr V_1$.}
\label{eq_diffusion_2}
\end{equation}
This equation, combined with the conservation of the diffusing species inside phase 1, $\nabla \cdot \mathbf J = \frac{\partial c}{\partial t}$, yields Fick's second law:
\begin{equation}
\frac{\partial c(\mathbf x, t)}{\partial t} = D \bigtriangleup c(\mathbf x, t), \mbox{ in $\mathscr V_1$.}
\label{eq_diffusion_3}
\end{equation}
If phase 2 are absorbing ``traps'' (rather than impenetrable obstacles as in the aforementioned effective diffusion problem), the boundary condition has to be changed. In the diffusion-controlled limit, i.e., when the reaction rate at the interface is infinite, we have the following boundary condition:\cite{torquato1991diffusion}
\begin{equation}
c(\mathbf x, t)=0\mbox{, on $\partial \mathscr V$.}
\end{equation}
If we also set the initial concentration to be uniform outside of traps:
\begin{equation}
c(\mathbf x, 0)=c_0\mbox{, in $\mathscr V_1$,}
\end{equation}
then we have the survival problem.
The ``survival probability,'' $p(t)$ is equal to the fraction of reactant not yet absorbed at time $t$:\cite{torquato1991diffusion, torquato2013random}
\begin{equation}
p(t)=\frac{\int_{\mathbb R^d} c(\mathbf x, t) d \mathbf x}{ \int_{\mathbb R^d} c(\mathbf x, 0) d \mathbf x}.
\end{equation}
The mean survival time of the reactant is the zeroth moment of $p(t)$:\footnote{Note that the commonly used notation for the mean survival time is $\tau$ \cite{torquato1991diffusion, torquato2013random}. Here, we use $T_{mean}$ to avoid confusion with the order metric $\tau$.}
\begin{equation}
T_{mean}= \int_0^\infty p(t) dt.
\end{equation}
The survival probability can be decomposed as a sum of exponential functions:
\begin{equation}
p(t) = \sum_{n=1}^\infty I_n \exp(-t/T_n),
\end{equation}
where $I_n$ are coefficients and $T_n$ are relaxation times. The largest relaxation time is called ``principal relaxation time'' and by convention denoted $T_1$.  These quantities can be measured directly by NMR experiments since in NMR experiment of fluid-saturated porous media, proton magnetization decays mainly on the phase boundary.\cite{straley1987magnetic, banavar1987magnetic, mitra1992effects}

It is worth noting that although the above problems involve differential equations, $D_{e}$, $p(t)$, and $T_{mean}$ can actually be calculated much more efficiently by simulating Brownian motions using the so-called ``first-passage time'' technique. (See the Sec.~\ref{simulation} for details.) The effective diffusion coefficient can be found from the ratio of the mean square displacement of such Brownian particles and the time spent. The survival probability $p(t)$ is equal to the probability that a Brownian particle have never reached any trap at time $t$. The mean survival time, $T_{mean}$, can be calculated by integrating $p(t)$ but can also be calculated, more easily, by finding the average time needed for a particle to reach a trap the first time. It is also worth noting that while the effective diffusion coefficient is identically zero as long as the void phase is not percolating, $T_{mean}$ and $T_1$ are both positive until the spheres cover the entire space.\cite{torquato2013random}

\subsection{Geometrical and topological properties}
This paper also studies the following geometrical and topological properties that are intimately related to the aforementioned diffusion characteristics.

\subsubsection{Hyperuniformity and stealthiness in many-particle systems and two-phase media}

As we have explained earlier, a hyperuniform many-particle system is one in which the structure factor, Eq.~(\ref{structurefactor}), approaches zero in the $\mathbf k \to \mathbf 0$ limit. The name ``hyperuniform'' refers to an anomalous suppression of density fluctuations: Consider random placements of a spherical observation window of radius $R$ in a $d$-dimensional many-particle system. The number of points contained in such window, $N(R)$, is a random variable. 
For a uniform but not hyperuniform many-particle system (e.g., ideal gas without a gravity field), $\sigma^2_N(R)$ for large $R$ scales as $R^{d}$. For a hyperuniform system, $\sigma^2_N(R)$ for large $R$ grows more slowly than $R^{d}$.
It has been proved that the above-mentioned two conditions of hyperuniformity, $\lim_{\mathbf k \to \mathbf 0} S(\mathbf k) = 0$ and $\sigma^2_N(R)$ for large $R$ grows more slowly than $R^{d}$, are mathematically equivalent.\cite{torquato2003local} 

A similar definition exists for two-phase media.\cite{torquato2016hyperuniformity} One can compute the volume fraction of either phase inside a spherical observation window of radius $R$ and find its variance. For large $R$, this variance scales as $R^{-d}$ for typical (non-hyperuniform) random two-phase media and decreases faster than $R^{-d}$ for hyperuniform two-phase media. An equivalent condition for hyperuniformity is that $\lim_{\mathbf k \to \mathbf 0} \tilde \chi_{_V}(\mathbf k)=0$, where 
\begin{equation}
\chi_{_V}(\mathbf k)=\frac{1}{v_F} |\mathcal J(\mathbf k)|^2
\label{chi_v}
\end{equation} 
is called the ``spectral density'' and $\mathcal J(\mathbf k)$ is the Fourier transform of $\mathcal I(\mathbf x)-\phi_1$.\cite{torquato1999exact}

Stealthy hyperuniform many-particle systems or two-phase media are subsets of hyperuniform many-particle systems or two-phase media in which $S(\mathbf k)$ or $\chi_{_V}(\mathbf k)$ is zero for a range of $\mathbf k$ vectors around the origin, i.e.,
\begin{equation}
S(\mathbf k) = 0 \mbox{ or } \chi_{_V}(\mathbf k)=0 \mbox{ for $0\le |\mathbf k| \le K$},
\end{equation}
where $K$ is some positive number. For the many-particle systems mentioned in this paper, the ground state of ``stealthy'' potentials are stealthy and hyperuniform while equilibrium hard-sphere systems and Poisson point process are neither stealthy nor hyperuniform.

\subsubsection{Packing and packing fraction}
When we decorate a point-particle configuration by replacing points with spheres of radius $a$, the whole collection of spheres is considered a ``sphere packing'' if each pair of point particles is separated by a distance of at least $2a$ (i.e., if the spheres do not overlap). The fraction of space occupied by the union of spheres, $\phi_2$, is called the packing fraction $\phi_p$. Of particular interest in this paper is the maximum packing radius $a^{max}_p$, which is half the minimum separation distance between two particles, and maximum packing fraction $\phi^{max}_p$, which is the volume fraction of phase 2 when $a=a^{max}_p$.

Why should we study the maximum packing fraction? One important reason is that when we decorate a point configuration and map it into a two-phase medium, if spheres do not overlap, then the spectral density $\tilde \chi_{_V}(\mathbf k)$ of the two-phase medium is proportional to the structure factor $S(\mathbf k)$ of the underlying point configuration:\cite{torquato2016disordered}
\begin{equation}
\tilde \chi_{_V}(\mathbf k)= \frac{\phi_2}{v_1(a)}\left(\frac{2\pi a}{|\mathbf k|}\right)^d J_{d/2}^2(|\mathbf k|a) S(\mathbf k) \mbox{ $(a \le a^{max}_p)$},
\label{chi_S}
\end{equation}
where $v_1(a)$ is the volume of a $d$-dimensional sphere of radius $a$ and $J_{d/2}(x)$ is the Bessel function of order $d/2$. Therefore, a decorated stealthy point configuration is a stealthy two-phase medium if $\phi_2<\phi^{max}_p$. When $\phi_2>\phi^{max}_p$, however, Eq.~(\ref{chi_S}) no longer holds and we will see in Sec.~\ref{Packing_Stealthy} that decorated systems are generally no longer stealthy or hyperuniform.

\subsubsection{Nearest-neighbor and pore-size functions}
Given a point-particle system, the void-exclusion probability $E_V(r)$ is the probability that a spherical cavity of radius $r$, centered at a random location, is empty of particles. A related quantity is $H_V(r)=-[\partial E_V(r)]/(\partial r)$, the probability density function of the distance to the nearest particle from a randomly chosen location. A different interpretation of $E_V$ is that if each point particle is replaced with a sphere of radius $a$, then $E_V(a)$ is the volume fraction of the space outside of the spheres, i.e., 
\begin{equation}
E_V(a)=\phi_1=1-\phi_2.
\label{Ev_phi}
\end{equation}
Since $H_V$ is the negative derivative of $E_V$, $H_V(a)$ is the specific surface $s$.\cite{torquato2013random}

Another quantity related to $E_V(r)$ is the scaled dimensionless quantizer error $\mathcal{G}$. 
For a point configuration with positions
${\bf r}_1,{\bf r}_2, \ldots$, a quantizer is a device that takes as an input a position $\mathbf x$ in $\mathbb R^d$ and outputs the nearest point $\mathbf r_i$ of the configuration to $\mathbf x$. Assuming $\mathbf x$ is uniformly distributed, one can define a mean square error, which can be obtained from $E_V(r)$ via the relation:\cite{torquato2010reformulation}
\begin{equation}
\label{QuantizerDefinition}
\mathcal{G}=\frac{2\rho^{\frac{2}{d}}}{d}\int _0 ^{\infty} r E_V(r) dr.
\end{equation}

Finally, two more related quantities can be defined for two-phase media. The pore-size cumulative distribution function, $F(\delta)$, is defined as the fraction of pore space (i.e., space covered by phase 1) which has a pore radius larger than $\delta$. The function $F(\delta)$ of our decorated system is trivially related to $E_V(r)$ of the underlying point-particle system:
\begin{equation}
F(\delta)=\frac{E_V(\delta+a)}{E_V(a)}.
\end{equation}
Moreover, the associated pore-size probability density function is given by
$P(\delta) = -[\partial F(\delta)]/(\partial \delta)$. This pore-size function at the origin is related to the specific surface, $s$, by
\begin{equation}
P(\delta=0)=\frac{s}{\phi_1}.
\end{equation}

It is interesting to note that the moments of $F(\delta)$ are related to the mean survival time and principle relaxation time via the following rigorous lower bounds \cite{torquato1991diffusion}:
\begin{equation}
T_{mean} \ge \frac{1}{D}\left( \int_0^\infty F(\delta) d\delta \right)^2,
\end{equation}
and
\begin{equation}
T_1 \ge \frac{2}{D}\int_0^\infty \delta F(\delta) d\delta.
\end{equation}
We see that $\mathcal{G}$ is proportional to the first moment of $F(\delta)$ in the $a \to 0$ limit and is therefore related to the principal relaxation time. 

\subsubsection{Order metric $\tau$} We will be studying the above properties for systems of varying degrees of order. Therefore, it is desirable to have a way to quantify such order. Moreover, since the underlying point-configurations we study include both stealthy ground states, which have long-range order, and equilibrium liquid hard-sphere systems, which have short-range order, we desire an order metric that reflects short-range order and long-range order equally well. A suitable choice is the order metric $\tau$, introduced in Ref.~\onlinecite{torquato2015ensemble} and defined as:
\begin{equation}
\tau=\frac{1}{D^d } \int_0^\infty [g_2(r)-1]^2 d\mathbf r = \frac{1}{(2\pi)^dD^d }\int_0^\infty [S(k)-1]^2 d\mathbf k,
\label{tau}
\end{equation}
where $D$ is some characteristic length scale, $g_2(r)$ is the pair correlation function,\cite{chandler1987introduction} $S(k)$ is the angular average of $S(\mathbf k)$, and the second equal sign can be proved by Parseval's theorem. In this paper, we simply let $D=1$ because we always rescale the configuration to make the number density unity.

\subsubsection{Percolation threshold and critical radius}
Since the effective diffusion coefficient is trivially zero when the void phase is topologically disconnected, it is important to quantify when the phases are connected. To do this, we will be considering the percolation properties of the systems.
As we specified earlier, we map point configurations into two-phase media by replacing each point with a sphere of radius $a$.
For phase 2, the critical or percolation radius, $a_{2c}$, is the minimum $a$ such that a connected part of phase 2 becomes infinite in size. The percolation volume fraction, $\phi_{2c}$, is the fraction of space occupied by the union of spheres of radius $a_{2c}$. 

We can define similar percolation characteristics of the void phase.\cite{rintoul2000precise, priour2014percolation} The percolation radius of the void phase, $a_{1c}$, is defined as the maximum $a$ such that there is still an infinite-sized connected part of phase 1. The percolation volume fraction, $\phi_{1c}$, is the volume fraction of phase 1 at radius $a_{1c}$. In two dimensions, it is very rare to have both phases percolating simultaneously (see Ref.~\onlinecite{sheng1982geometric} for such a rare example). In our case, $a_{1c}=a_{2c}$ and $\phi_{1c}=1-\phi_{2c}$. In three dimensions, however, both phases can simultaneously percolate, i.e., the two-phase system is bicontinuous. Indeed, this is the case for our 3D systems and hence we must compute $a_{1c}$ and $a_{2c}$ separately.

\section{simulation details}
\label{simulation}

\subsection{Generating entropically favored stealthy ground states}
\label{GeneratingStealthy}

We generate entropically favored ground states of stealthy potentials using the same protocol as our previous work.\cite{zhang2015ground} This protocol involves performing molecular dynamics (MD) simulations at a very low temperature ($\beta = 5\e{5}$ in 2D and $\beta=1\e{6}$ in 3D in dimensionless units), taking snapshots periodically, and performing a local energy minimization starting from each snapshot. Because the MD temperature is sufficiently low, the snapshots before energy minimization are already very close to ground states. Therefore, the ground states produced by the subsequent energy minimization closely follow the canonical distribution in the zero-temperature limit. We generate 20,000 configurations per $\chi$ value, same as Ref.~\onlinecite{zhang2015ground}. The only two differences between this work and our previous work \cite{zhang2015ground} are (1) system sizes are different (see Appendix~\ref{number} for our choice of system sizes and the justification), and (2) each configuration is rescaled to unit number density (in order to ensure a fair comparison).



\subsection{Generating equilibrium disordered hard-sphere systems}
We also generate equilibrium disordered hard-sphere systems via standard Monte-Carlo techniques in order to compare their statistics with entropically favored stealthy ground states' statistics. Depending on the packing fraction $\phi$, an equilibrium hard-sphere system can be disordered (liquid-like) or crystalline. Disordered equilibrium hard-sphere system exists for $0<\phi<0.69$ in 2D and $0<\phi<0.49$ in 3D.\cite{torquato2013random} Therefore, the packing fraction we used include $\phi=0.05$, 0.1, 0.15, 0.2, 0.25, 0.3, 0.35, 0.4, 0.45, 0.5, 0.55, 0.6, 0.65, and 0.68 in 2D and $\phi=0.05$, 0.1, 0.15, 0.2, 0.25, 0.3, 0.35, 0.4, 0.45, and 0.48 in 3D. For each $\phi$ in each dimension, we generate equilibrium hard-sphere systems with $N=100$, 300, and 500 particles. In each case, the system was first equilibrated with $3\e{6}N$ trial moves. After that, we sample a configuration every $300N$ trial moves until we obtain 20,000 configurations. Similar to the stealthy ground states, we keep the number density $\rho=1$. Therefore, we adjust sphere radius to attain a desired packing fraction.

\subsection{Calculating survival probability, mean survival time, and principal relaxation time}
Because the method we used to calculate the effective diffusion coefficient is an extension of the method to calculate survival probability and mean survival time, we will explain the latter method first.
The survival probability $p(t)$ and mean survival time $T_{mean}$ can be calculated by simulating particles undergoing Brownian motions. 

The Brownian motion can be simulated very efficiently using the first-passage-time technique.\cite{torquato1989efficient} The key idea of this technique is that for a Brownian particle at a particular location, let ${\mathscr R}$ be the distance between it and the closest phase boundary. Construct a sphere centered at the particle with radius ${\mathscr R}$ (which is called a first-passage-time sphere). Let $t_{\mathscr R}$ be the time needed for the particle to reach the surface of such sphere for the first time, the distribution of $t_{\mathscr R}$ can be calculated analytically. In 3D, the cumulative distribution function (CDF) of $t_{\mathscr R}$ is\cite{torquato1989efficient}
\begin{equation}
F(t_{\mathscr R}) = 1 + 2 \sum_{m=1}^\infty (-1)^m \exp\left( - \frac{Dm^2\pi^2 t_{\mathscr R}}{{\mathscr R}^2} \right).
\label{CDF_3D}
\end{equation}
In 2D, Ref.~\onlinecite{torquato1989efficient} did not provide the distribution of $t_R$. Here we find the following explicit 2D expression for $t_R$:
\begin{equation}
F(t_{\mathscr R}) = 1-2\sum_{m=1}^\infty \frac{\exp(-D w_m^2 t_{\mathscr R}/{\mathscr R}^2)}{w_m J_1(w_m)},
\label{CDF_2D}
\end{equation}
where $J_n(x)$ is the Bessel function of order $n$, and $w_n$ is the $n$th root of $J_0(x)$. The mean of $t_{\mathscr R}$, in any dimension, is simply ${\mathscr R}^2/2dD$.

Therefore, the Brownian motion inside the first-passage-time sphere does not need to be simulated in detail. One simply moves the particle to a random location on the surface of such sphere, and increase the time by a certain amount, as detailed below. When calculating the mean survival time $T_{mean}$, the time increment can simply be ${\mathscr R}^2/2dD$, the mean of $t_{\mathscr R}$. When calculating $p(t)$, however, the time increment has to be a random number drawn from the distributions given in Eq.~(\ref{CDF_3D}) or (\ref{CDF_2D}). The process of finding ${\mathscr R}$, moving the particle, and increasing the time is repeated until the Brownian particle gets very close ($\ee{-5} a$) to a trap, at which time the Brownian particle is deemed trapped. In our implementation, Eqs.~(\ref{CDF_3D})-(\ref{CDF_2D}) are pre-computed and tabulated to accelerate the simulation. For each configuration, we simulate 10 Brownian trajectories to calculate $T_{mean}$ and 1000 trajectories to calculate $p(t)$. When calculating $p(t)$, each trajectory is additionally sampled 100 times, with different random time increments drawn from distributions (\ref{CDF_3D})-(\ref{CDF_2D}).

After calculating $p(t)$, we calculate the principal relaxation time $T_1$ by fitting $p(t)$ in the range $\ee{-5}<p(t)<\ee{-3}$ to the asymptotic equation
\begin{equation}
\ln[p(t)] \approx c + t/T_1,
\end{equation}
where $c$ and $T_1$ are fitting parameters.

\subsection{Calculating effective diffusion coefficient}
The effective diffusion coefficient $D_e$ can also be calculated using first-passage-time techniques. \cite{kim1990determination, kim1992effective, torquato1999effective} In this case, however, the Brownian particle cannot be deemed trapped when it is sufficiently close to the phase boundary because phase 2 is now non-absorbing obstacles rather than absorbing traps. Instead, we construct a first-passage-time sphere of radius ${\mathscr R} =\ee{-2}$, find a random place on the surface of the first-passage-time sphere that is outside of the obstacle phase, and move the Brownian particle to that random place. 
Although this first-passage-time sphere contains two phases, 
the mean time taken for the Brownian particle to reach such surface could still be computed analytically and was given in Ref.~\onlinecite{kim1992effective}:
\begin{equation}
t_{\mathscr R} = \frac{{\mathscr R}^2(1+v_2/v_1)}{2d},
\end{equation}
where $v_2/v_1$ is the volume of the obstacle phase divided by the volume of the conducting phase inside the first-passage-time sphere and can be found analytically.

The process of constructing a first-passage-time sphere and moving the point particle is repeated to form a Brownian trajectory. In the infinite-time limit, the effective diffusion coefficient is given by:\cite{kim1992effective}
\begin{equation}
D_e=\lim_{t \to \infty} \frac{<|\mathbf R(t)|^2>}{2dt},
\label{eq_De}
\end{equation}
where $<|\mathbf R(t)|^2>$ is the mean-squared displacement of a Brownian particle at time $t$. In practice, in a finite-time simulation, one should only consider the time regime in which
the mean square displacement
is strongly linear in time, since
for sufficiently early times 
the mean square displacement
is either ballistic or grows
faster than linear in time.\cite{torquato1999effective}
We find $D_e$ by fitting $<|\mathbf R(t)|^2>$ versus $t$ and extracting the slope of the line after some sufficiently large dimensionless time. We define the unit of time to be
\begin{equation}
t^*=\frac{1}{\rho^{2/d}D},
\end{equation}
and set both $\rho$ and $D$ to be unity.
The point in time in which $<|\mathbf R(t)|^2>$ first becomes a strongly linear function occurs when the Brownian particle sufficiently samples the two-phase system such that it can be viewed effectively as Brownian motion in a homogeneous medium. For the microstructures that we considered here, we find that the linear regime occurs in the dimensionless time interval $40<t<100$, i.e., we determine $D_e$ from the linear relationship
\begin{equation}
<|\mathbf R(t)|^2>=(2dD_e)t +c, \mbox{ $40<t<100$}.
\label{eq_De_fit}
\end{equation}

\begin{figure}[h]
\includegraphics[width=0.3\textwidth]{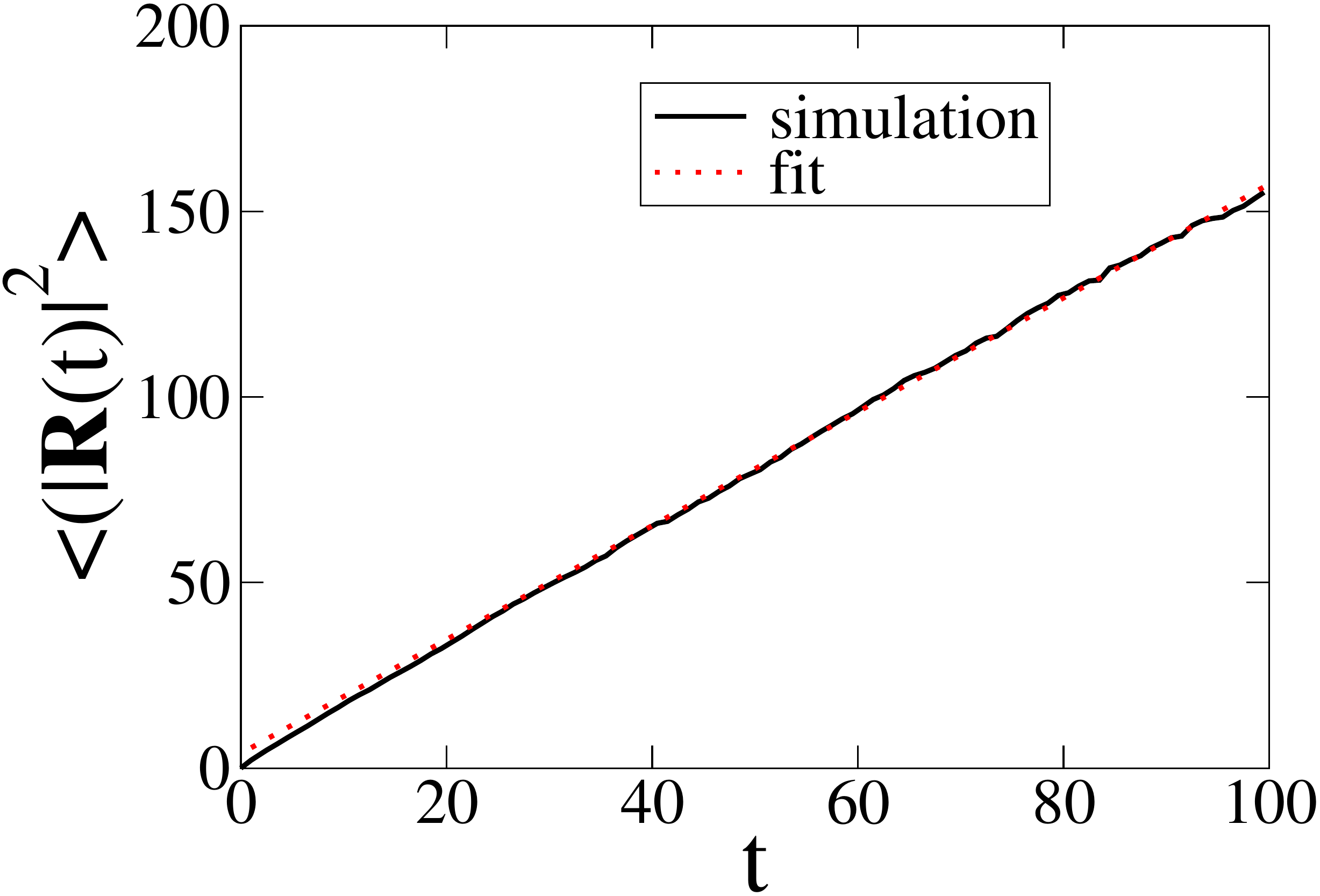}
\includegraphics[width=0.3\textwidth]{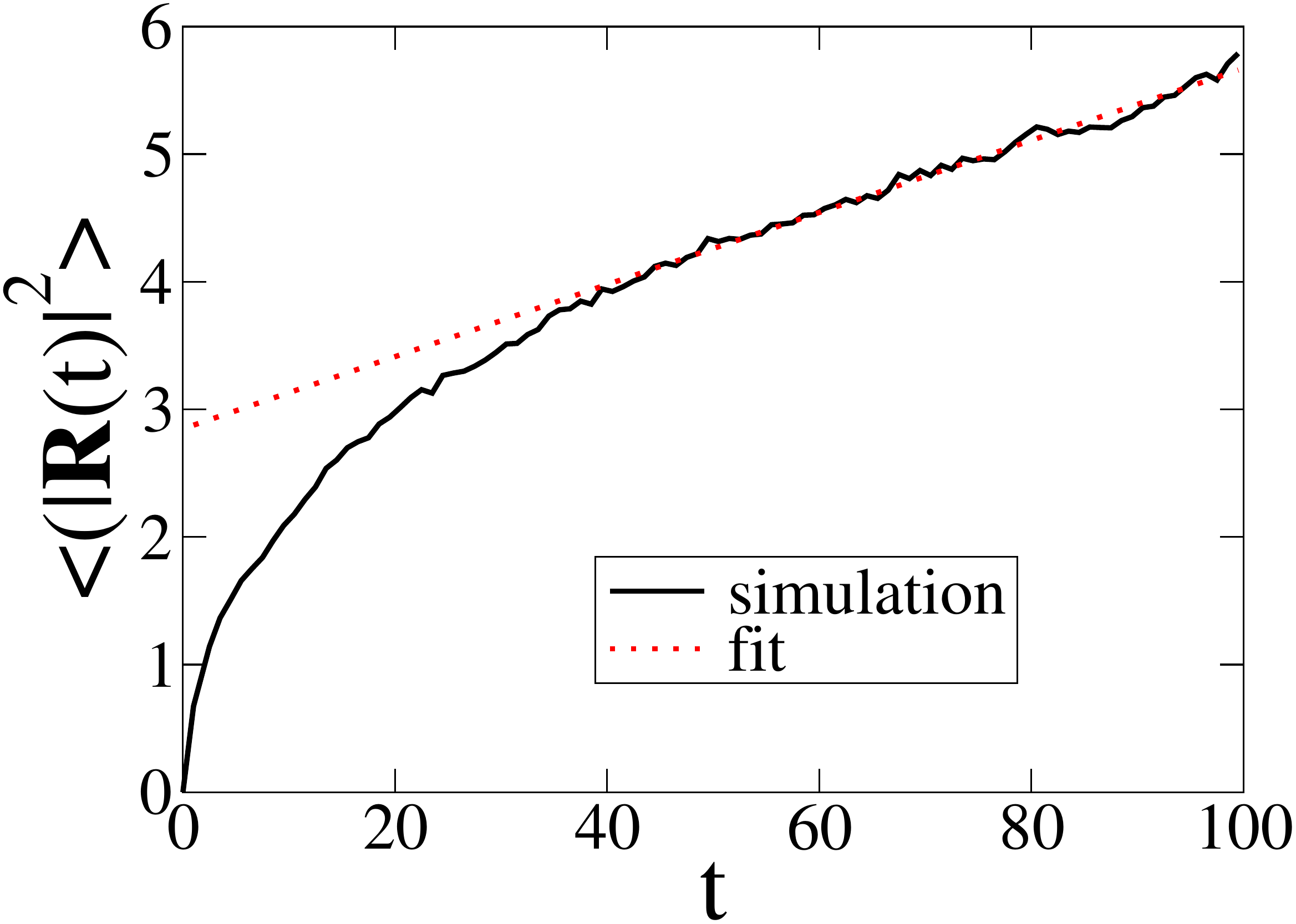}
\includegraphics[width=0.3\textwidth]{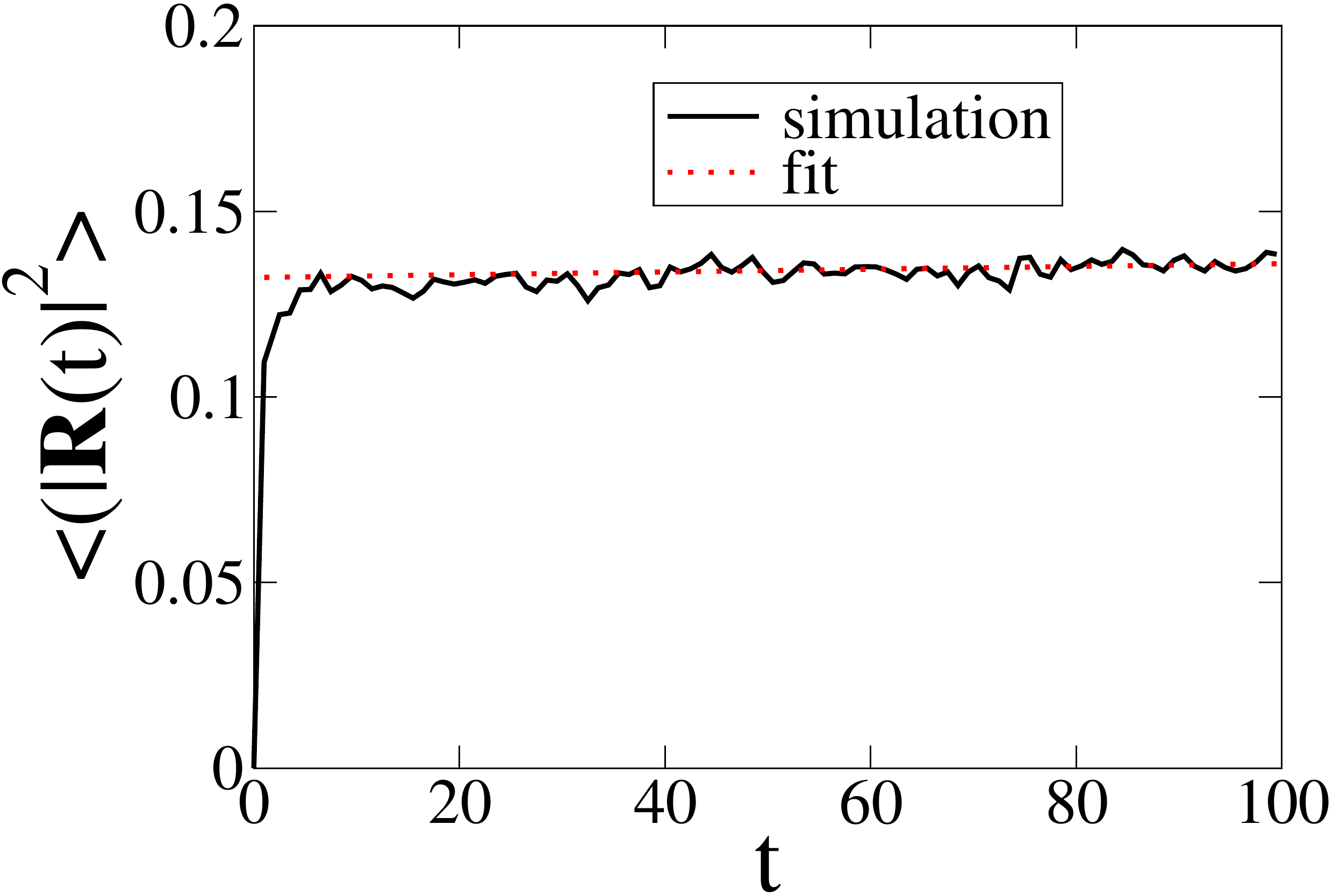}
\caption{The mean square displacement of Brownian particles, averaged over 20,000 configurations, $<|\mathbf R(t)|^2>$, versus time, $t$, for a three-dimensional system at $\chi=0.1333\cdots$ with obstacle radii $a=0.7$ (top), 0.8 (middle), and 0.9 (bottom). For this particular system, the percolation threshold of the void phase is $a_{1c}=0.80$, and hence $D_e$ must vanish for larger values of $a$.}
\label{r2vt}
\end{figure}

To get a sense of the possible behaviors of the mean square displacements as a function of time, we show examples in Fig.~\ref{r2vt} at several values of $a$ for a three-dimensional system at $\chi=0.1333\cdots$ and indicate the linear fit in each case. This fitting procedure works especially well near percolation, which is the most difficult regime to simulate. For this particular system, the void phase stops percolating at obstacle radius $a_{1c}=0.80$. Figure~\ref{r2vt} shows that for $a=0.7<a_{1c}$, $<|\mathbf R(t)|^2>$ is linear with $t$. For $a=0.8=a_{1c}$, only a fraction of configurations still have a percolating void phase, and our fitting procedure was able to distinguish the initial uprise in $<|\mathbf R(t)|^2>$ (contributions mainly from Brownian particles moving inside a ``cage'', i.e., a disconnected part of the void phase) from the steady increase in $<|\mathbf R(t)|^2>$ (contributions from Brownian particles that are in a percolating part of the void phase). For $a=0.9>a_{1c}$, all Brownian particles are caged, and the fit has a virtually zero slope (and therefore produces a virtually zero $D_e$). 

\begin{figure}[h]
\includegraphics[width=0.5\textwidth]{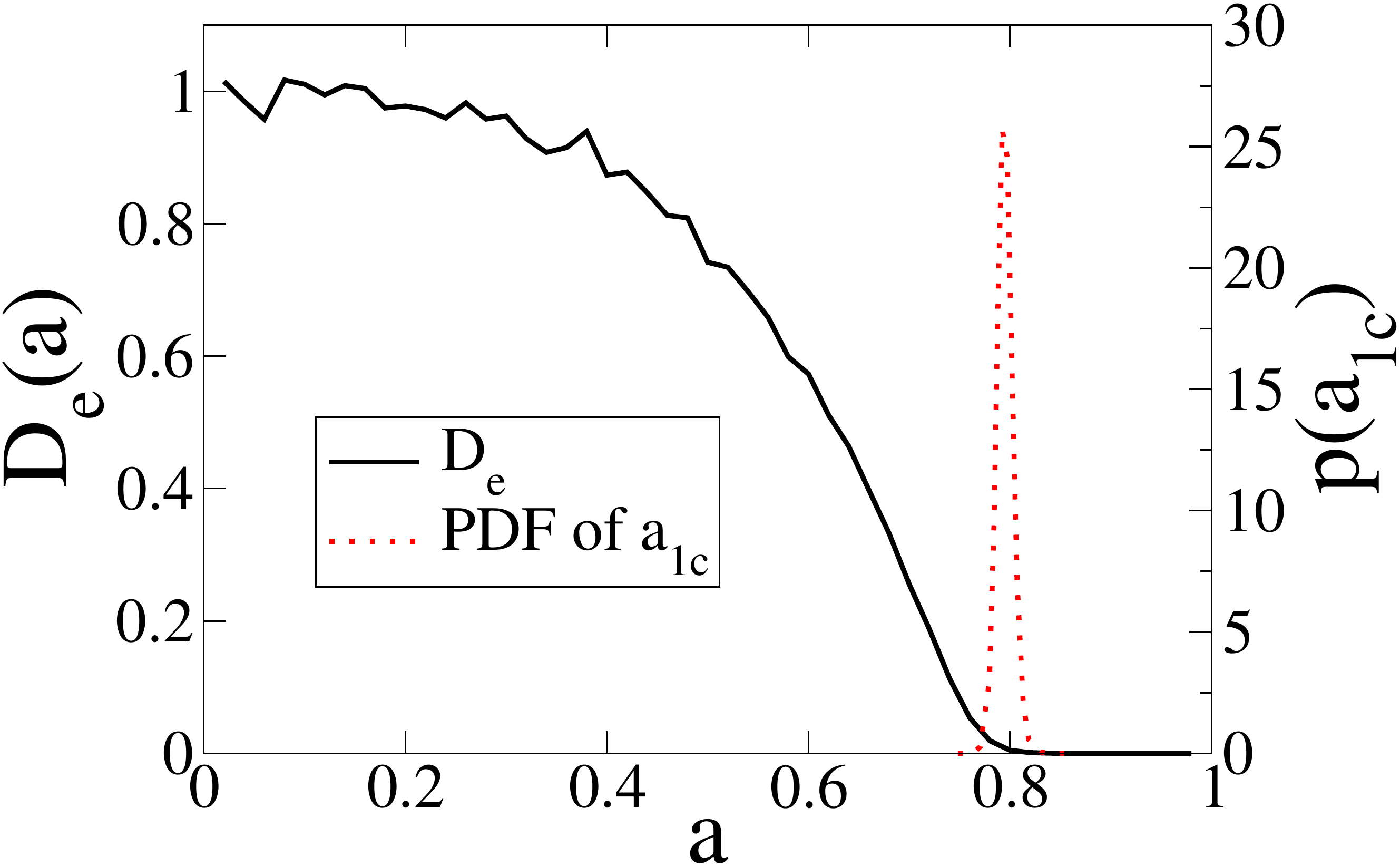}
\caption{Comparing the calculated effective diffusion coefficient, $D_e$, with the probability density function (PDF) of the void-phase percolation threshold, $p(a_{1c})$, for three dimensions, $\chi=0.1333$. The calculated $D_e$ becomes zero when the void phase stops percolating.}
\label{De_withPerco}
\end{figure}

We simulate 1 Brownian trajectory per configuration to calculate $D_e$. 
In Fig.~\ref{De_withPerco}, we compare the computed $D_e$ with the distribution of the void-phase percolation threshold and find  that $D_e$ becomes zero right after all configurations stop percolating.
The fact that our measured $D_e$ diminishes to zero at the percolation threshold indicates that our choice of the fitting range in Eq.~(\ref{eq_De_fit}) is appropriate.

\subsection{Calculating percolation thresholds}

Generally speaking, the precise calculation of the percolation threshold of disordered systems require very large system sizes. For example, to accurately determine the percolation threshold of 3D  fully penetrable spheres, Ref.~\onlinecite{lorenz2001precise} employed systems of up to $N=7\e{8}$ particles. The whole system is divided into smaller cubes and the content particles in each cube is generated only when such cube is being probed.

Unfortunately, our protocol of low-temperature MD and a subsequent energy minimization does not allow us to save time by only generating required parts of the configuration. Moreover, in order to accurately follow the canonical-ensemble distribution at zero-temperature limit, the MD temperature has to be so low such that many ($7.5\e{7}$) time steps are required to produce a sufficiently long trajectory. The requirement of a very large number of time steps forces us to further sacrifice system size. As a result, our system is limited to several hundred particles. Therefore, accurate determination of the percolation threshold is extremely challenging. Thus we experimented with two advanced algorithms to minimize finite-size effect in order to obtain relatively accurate results. We will first explain how to use these two methods to determine the percolation threshold for the particle phase, and then describe the generalizations to the void phase.

One of them, which we call ``$P_1$ maximum method,'' is described in Ref.~\onlinecite{newman2000efficient}. Starting from a random particle in a configuration, one randomly chooses two of its periodic images in two different directions. The quantity $P_1$ (denoted as $R_\infty^{(1)}$ in Ref.~\onlinecite{newman2000efficient}) is defined as the probability that this particle is connected to one of the chosen periodic images but not the other. At the percolation threshold, $P_1$ attains its maximum. Therefore, one can numerically find $P_1$ as a function of sphere radius $a$ and find its maximum in order to find the percolation threshold. In our implementation, we calculate $P_1(a)$ for various $a$'s starting from $a=0$, with increment $\delta a=0.001$, until $P_1(a)$ develops a peak and then returns to zero. We then select all data points such that $P_1(a)>0.9 \cdot P_1^{max}$, where $P_1^{max}$ is the maximum of $P_1(a)$, and perform a quadratic fit of the selected data points. The maximum of the fitted function gives the percolation radius $a_{2c}$. 

Ref.~\onlinecite{newman2000efficient} measures the percolation radii $a_{2c}$ using several different system sizes and then extrapolates to the infinite-system-size limit. However, when we perform the same fitting procedure using different system sizes, we did not find a clear trend: In each dimension, for some $\chi$ values larger systems produces larger $a_{2c}$ while for other $\chi$ values larger systems produces smaller $a_{2c}$. Moreover, the extrapolated $a_{2c}$, as a function of $\chi$, is not as smooth as the un-extrapolated one. We therefore conclude that random noise is probably more important than finite-size effect in this case and extrapolation is not proper. Thus, we will simply use $a_{2c}$ of our largest system as an estimate of the infinite-system-size $a_{2c}$.

After finding the percolation radius $a_{2c}$, we determine the percolation volume fraction $\phi_{2c}$. One could have simply read this quantity from a plot of the quantity $E_V(r)$, since $\phi_{2c}=1-E_V(a_{2c})$. However, we decide to use a somewhat more accurate method: we divide the whole simulation box into 12000$\times$12000 pixels (in 2D) or  1200$\times$1200$\times$1200 voxels (in 3D) and find out if the center of each pixel or voxel is inside any sphere of radius $a_{c}$. We then count the number of pixels or voxels that are centered inside spheres to find out the volume fraction. From our experience, this procedure gives us a four-significant-figures precision in $\phi_{c}$.

The other method we employed, which we call ``$M_2$ intersection method,'' is introduced in Ref.~\onlinecite{ziff2016percolation}. At a given radius $a$, define $s_{max}$ to be the size of the largest cluster, $M_2$ (denoted as $R_2$ in Ref.~\onlinecite{ziff2016percolation}) is defined as:
\begin{equation}
M_2= \frac{\langle s_{max}^2 \rangle -\langle s_{max} \rangle ^2}{\langle s_{max} \rangle ^2},
\end{equation}
where $\langle \cdots \rangle$ denotes an ensemble average. As Ref.~\onlinecite{ziff2016percolation} shows, at the percolation threshold, $M_2$ is the same for different system sizes. Therefore, one can compute $M_2$ as a function of $a$, and find the intersection of $M_2(a)$ for different system sizes to find $a_{2c}$. Following Ref.~\onlinecite{ziff2016percolation}, we use three different $N$'s for each $\chi$ value, and perform an extrapolation to find $a_{2c}$ in the infinite-$N$ limit. After that, we use the same procedure discussed in the previous paragraph to calculate $\phi_{2c}$ from $a_{2c}$.

\begin{figure}[h]
\includegraphics[width=0.45\textwidth]{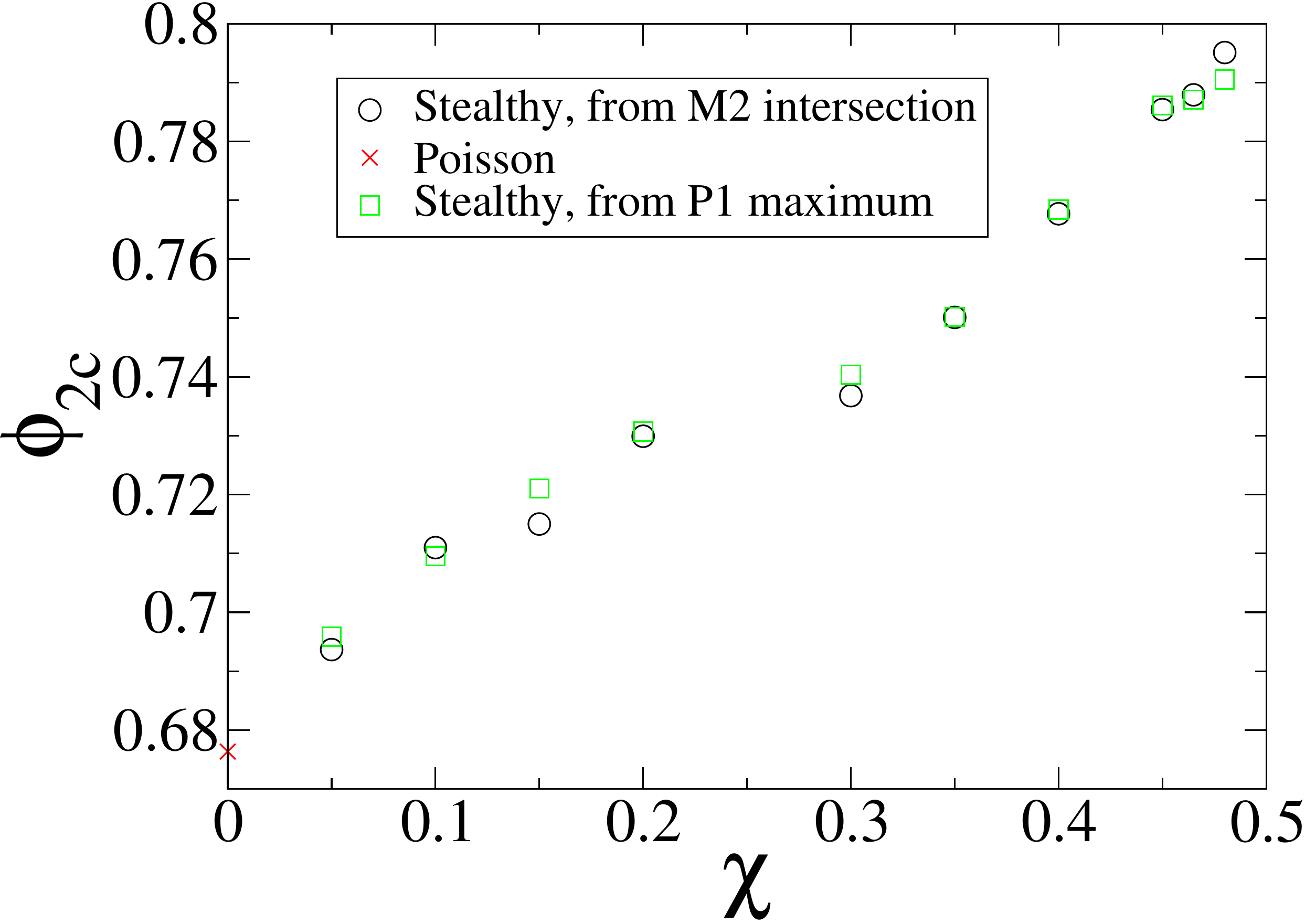}
\includegraphics[width=0.45\textwidth]{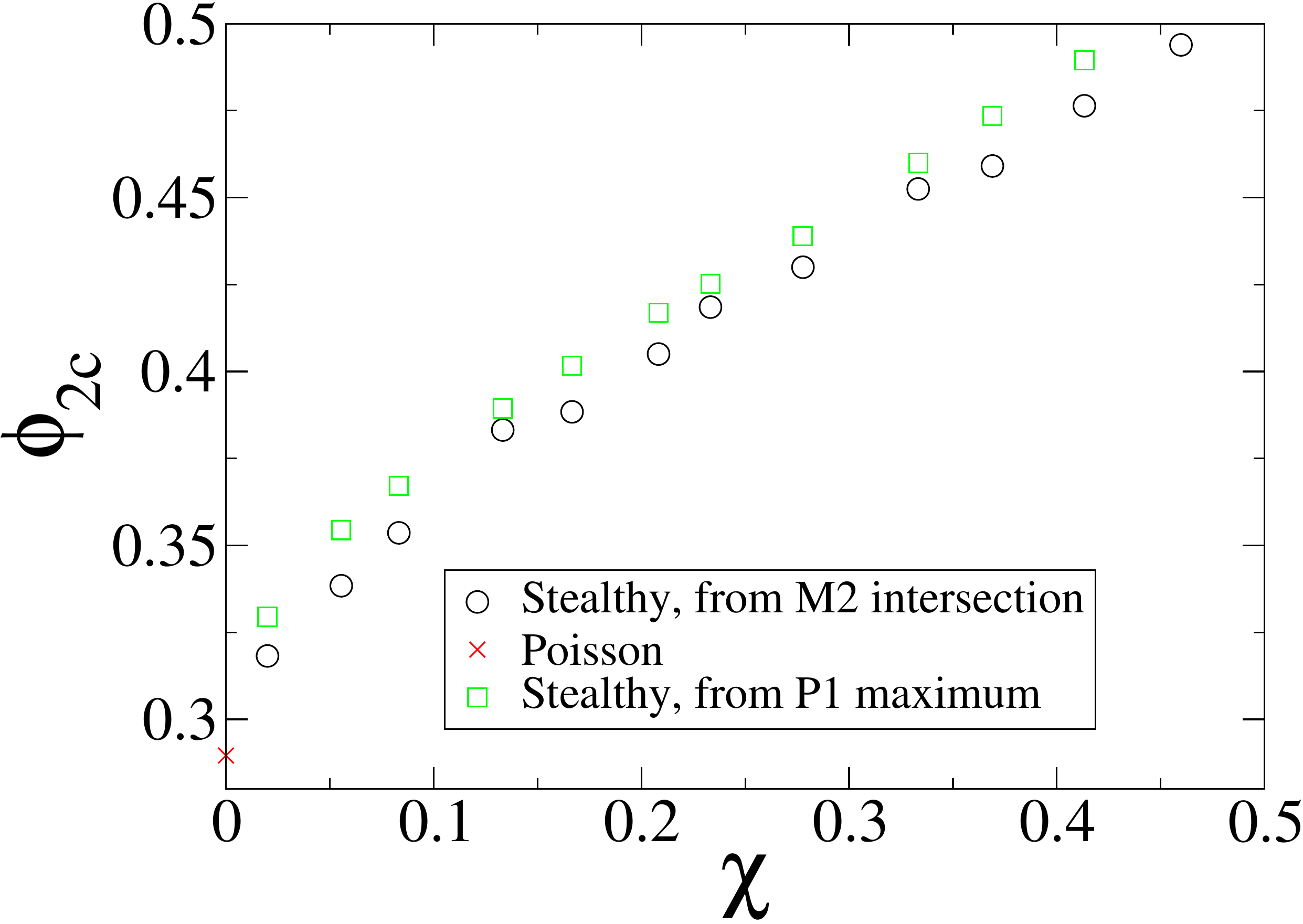}
\caption{Particle-phase percolation volume fraction $\phi_{2c}$ of entropically favored stealthy ground states at different $\chi$'s in 2D (top) and 3D (bottom).}
\label{perco}
\end{figure}

We have used both methods to calculate the percolation volume fraction $\phi_{2c}$ of decorated stealthy ground states at various $\chi$'s in 2D and 3D. They are presented in Fig.~\ref{perco}. Figure~\ref{perco} also presents $\phi_{2c}$ for decorated Poisson point processes obtained from Refs.~\onlinecite{lorenz2001precise} and \onlinecite{quintanilla2000efficient}, which are $\phi_{2c}=0.676339$ in 2D and $\phi_{2c}=0.289573$ in 3D. These results can be used as benchmarks since Poisson point processes are geometrically equivalent to entropically favored stealthy ground states at $\chi=0$. We see that in 2D, while both methods give results that approaches the Poisson value very well in the $\chi\to 0$ limit, the $P_1$ maximum method produces much smoother results. In 3D, however, although both methods produce relatively smooth results, only results from the $M_2$ intersection method approaches the Poisson value very well in the $\chi\to 0$ limit. Therefore, we decide to choose the $P_1$ maximum method in 2D and the $M_2$ intersection method in 3D for the rest of the paper. It is interesting to note that to our knowledge, the $P_1$ maximum method has been demonstrated to work well in 2D \cite{newman2000efficient} but not in 3D, while the $M_2$ intersection method has been demonstrated to work well in 3D \cite{ziff2016percolation} but not in 2D. It is possible that these two methods are just more suited to their respective dimensions.

\begin{figure}
\includegraphics[width=0.45\textwidth]{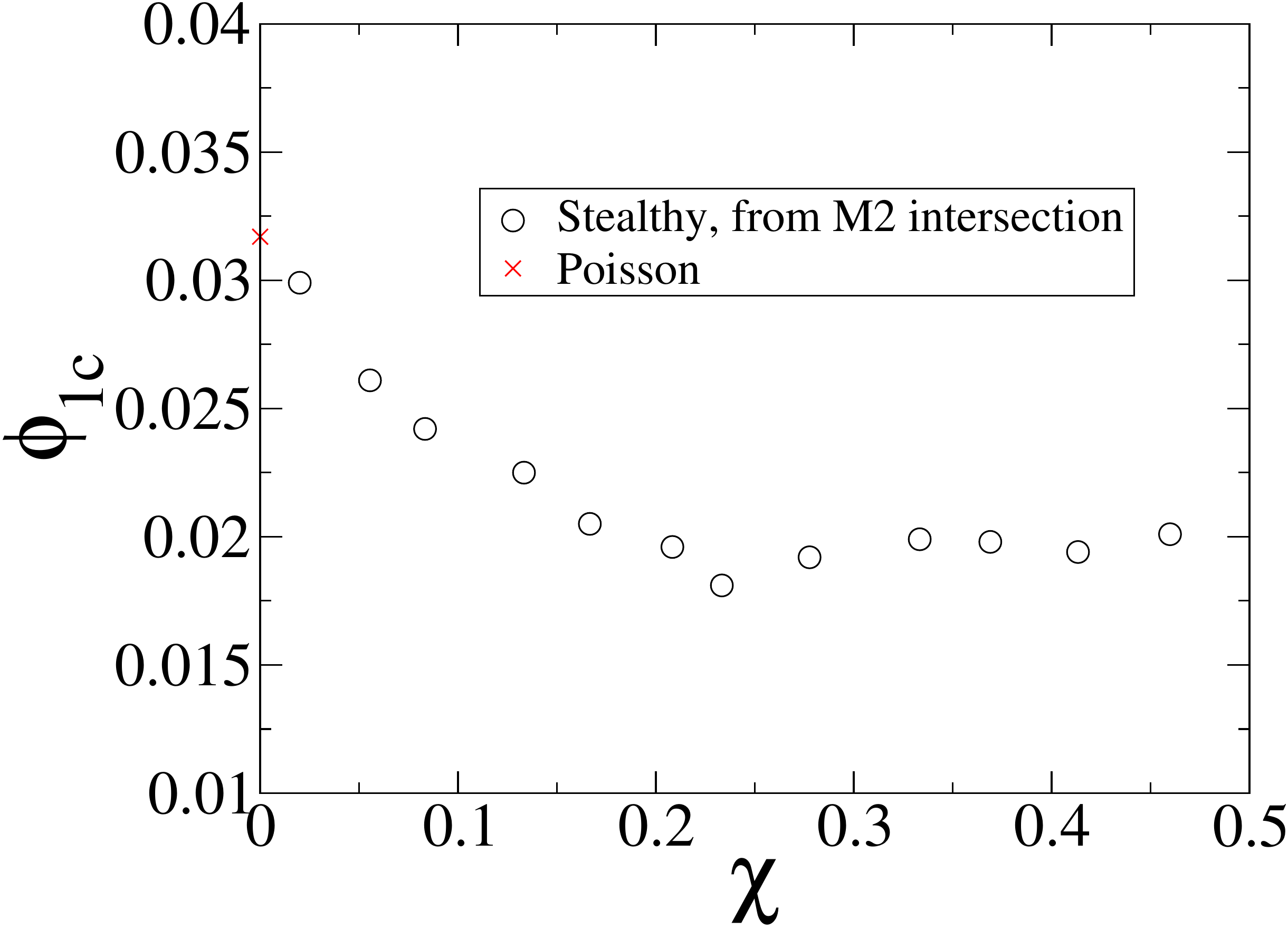}
\caption{Void-phase percolation volume fraction $\phi_{1c}$ of entropically favored stealthy ground states at different $\chi$'s in 3D.}
\label{perco1}
\end{figure}

Besides the percolation threshold of the spheres, we also study the percolation threshold of the void phase. In two dimensions, the percolation radius of the void phase, $a_{1c}$, is equal to the percolation radius of the spheres, $a_{2c}$. In three dimensions, however, $a_{1c}$ has to be calculated separately. We compute $a_{1c}$ in three dimensions by performing a Voronoi tessellation  of each configuration, and then computing $M_2$ of the Voronoi vertices. As in the particle-phase case, the intersection of $M_2(a)$ at different system sizes gives $a_{1c}$. Similar to the particle-phase case, $a_{1c}$ can then be converted to $\phi_{1c}$ by digitization, the result of which is presented in Fig.~\ref{perco1}. Similar to the particle-phase case, we compare $\phi_{1c}$ for our systems with that for the decorated Poisson point processes obtained from Ref.~\onlinecite{priour2014percolation}, $\phi_{1c}=0.0317$. Combining the $\phi_{2c}$ and $\phi_{1c}$ results, we see that as $\chi$ increases from 0 to 0.46, the $\phi_{2}$ range for bicontinuity moves upwards, from $0.290<\phi_{2}<0.997$ to $0.494<\phi_{2}<0.998$, respectively.

\subsection{Calculating $E_V(r)$, $\mathcal{G}$, and $\tau$}

The quantities $E_V(r)$ and $\mathcal{G}$ are calculated by first computing $H_V(r)$. For each configuration of $N$ point particles, $100N$ random locations in 2D or $10N$ random locations in 3D are generated in the simulation box. For each location, the distance from it to its nearest particle is found. These distances are then binned to yield $H_V(r)$. We then integrate $H_V(r)$ using trapezoidal rule to find $E_V(r)$. The quantizer error $\mathcal{G}$ is obtained by another integration of $rE_V(r)$, using trapezoidal rule, as Eq.~(\ref{QuantizerDefinition}) shows. The numerically obtained $H_V(r)$ always have compact support, and thus the above-mentioned integrations does not need to be truncated.

The order metric $\tau$ can be computed from either $g_2(r)$ or $S(k)$, as Eq.~(\ref{tau}) shows. We have tried both approaches. The real-space integration in Eq.~(\ref{tau}) is truncated at half the simulation box side length and the reciprocal space integration in Eq.~(\ref{tau}) is truncated at $6K$, where $K$ is the cutoff of the stealthy potential (as detailed in Sec.~\ref{subsection_StealthyPotential}).

\section{Results}
\label{result}

\begin{figure}[H]
\begin{center}
\includegraphics[width=0.48\textwidth]{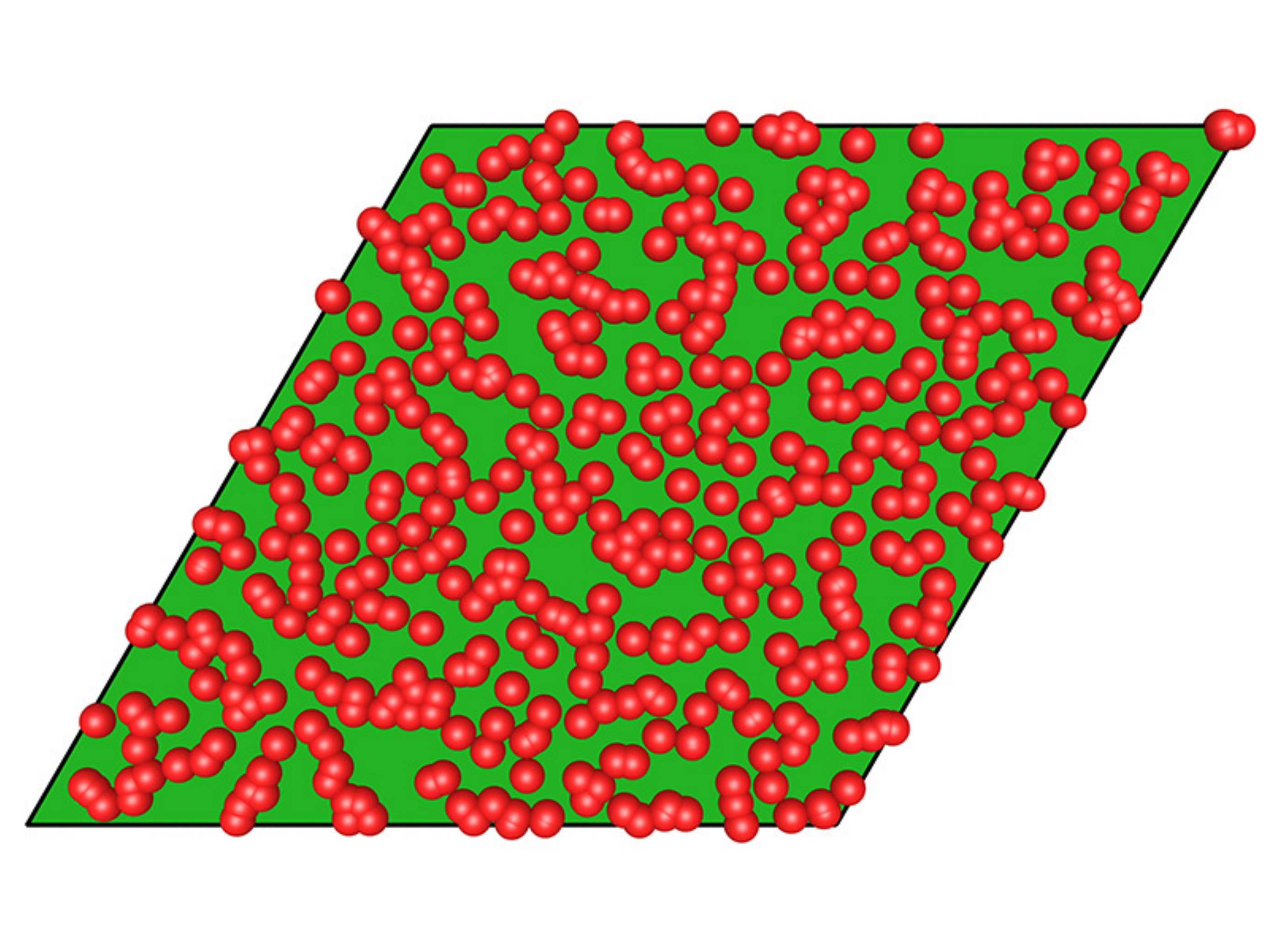}
\includegraphics[width=0.48\textwidth]{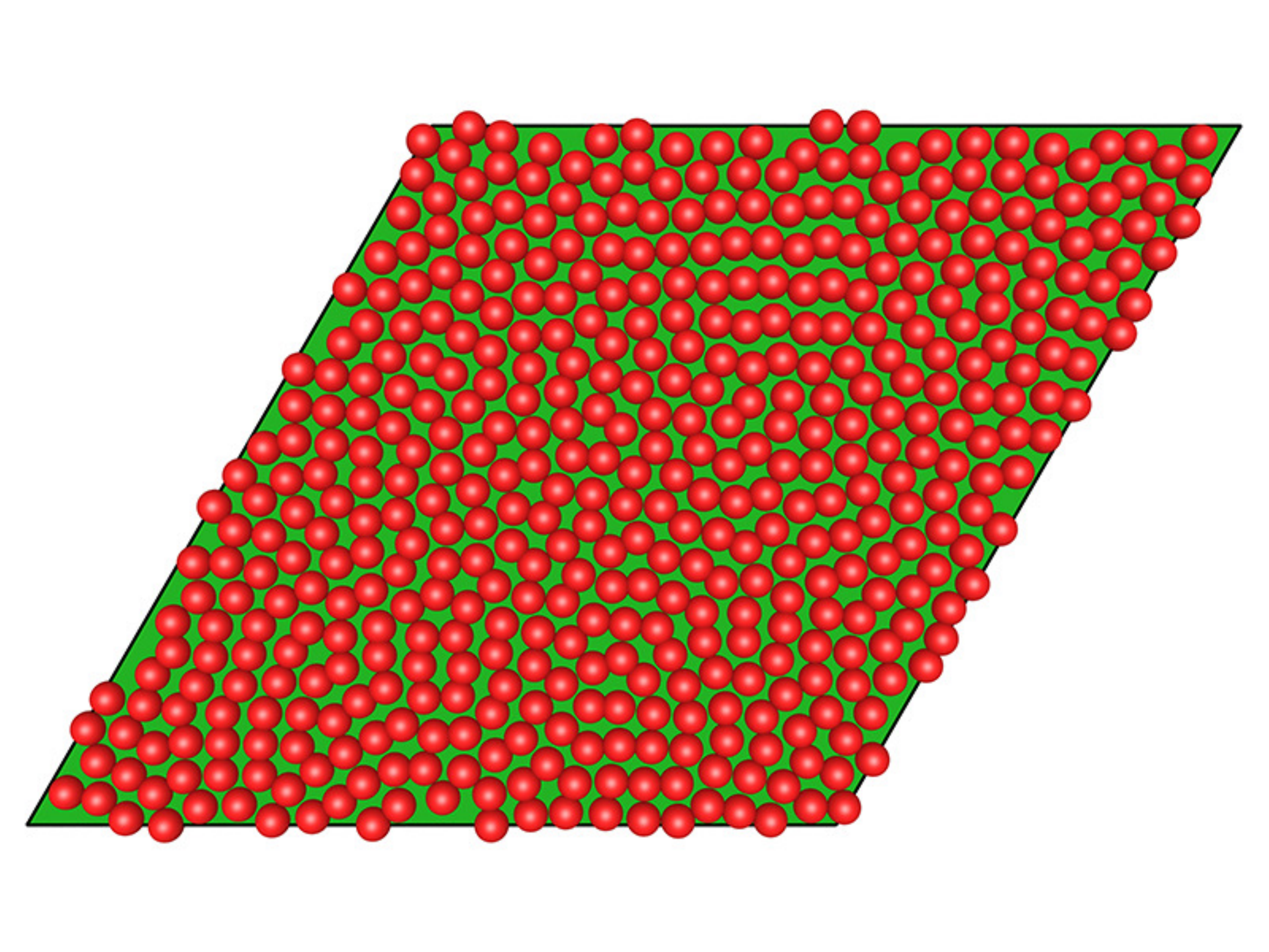}
\end{center}
\caption{Decorated stealthy ground states in two dimensions at $\chi=0.05$ (top) and $\chi=0.48$ (bottom), at $a=0.5$. The void phase is marked green. }
\label{Configurations_2}
\end{figure}

\begin{figure}[H]
\begin{center}
\includegraphics[width=0.48\textwidth]{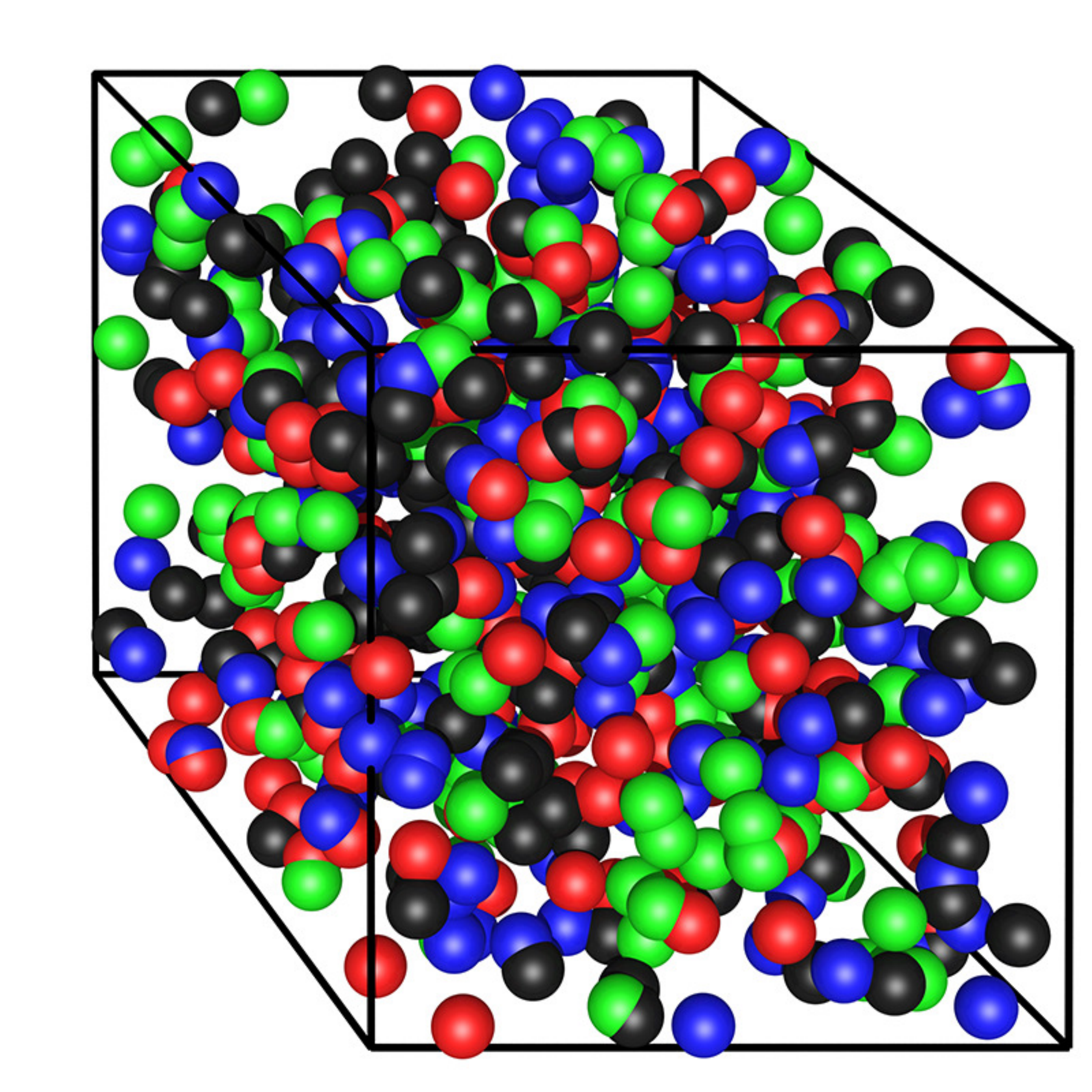}
\includegraphics[width=0.48\textwidth]{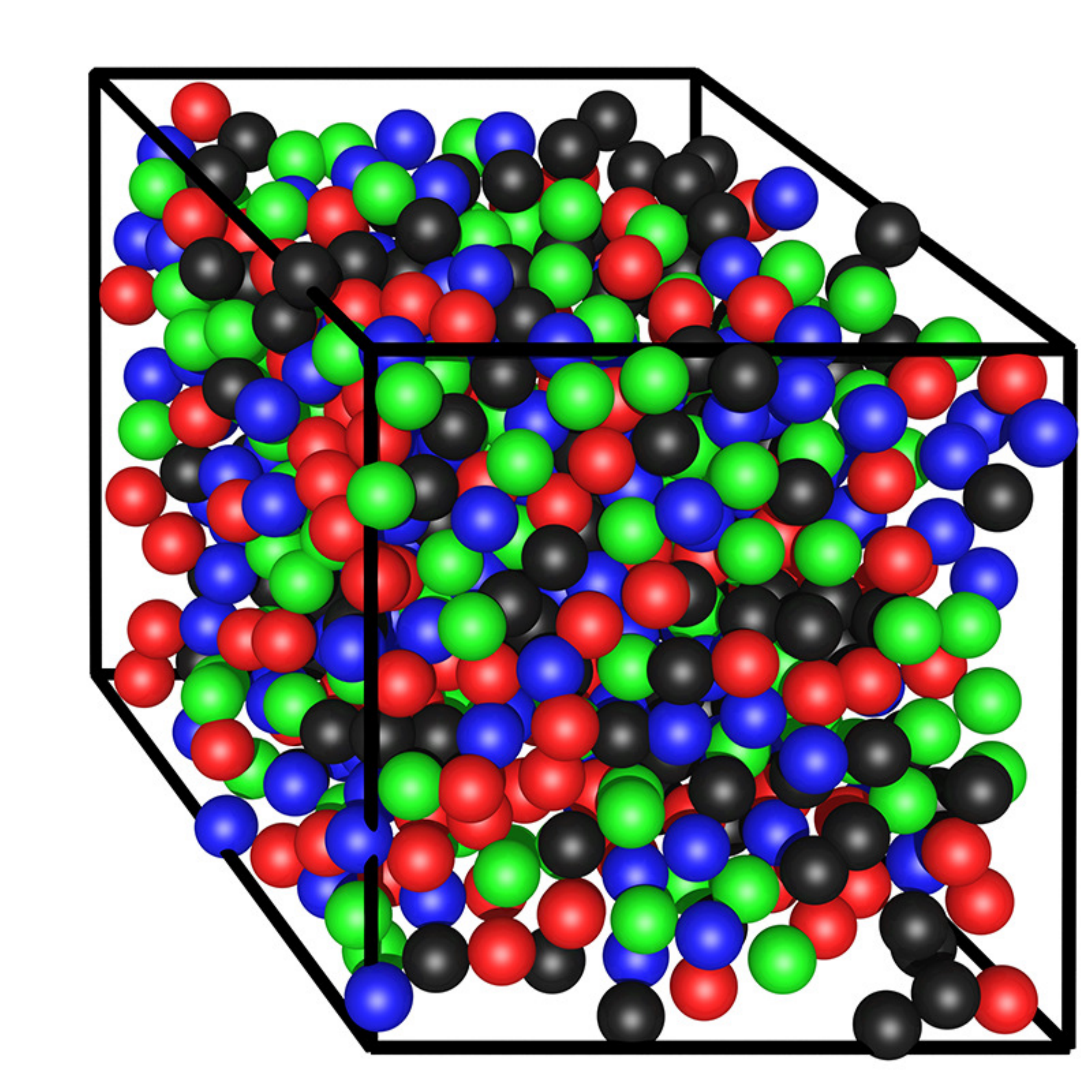}
\end{center}
\caption{Decorated stealthy ground states in three dimensions at $\chi=0.02$ (top) and $\chi=0.4598...$ (bottom), at $a=0.5$. Each sphere is randomly assigned to one of four colors in order to improve visual clarity.}
\label{Configurations_31}
\end{figure}

\begin{figure}[H]
\begin{center}
\includegraphics[width=0.48\textwidth]{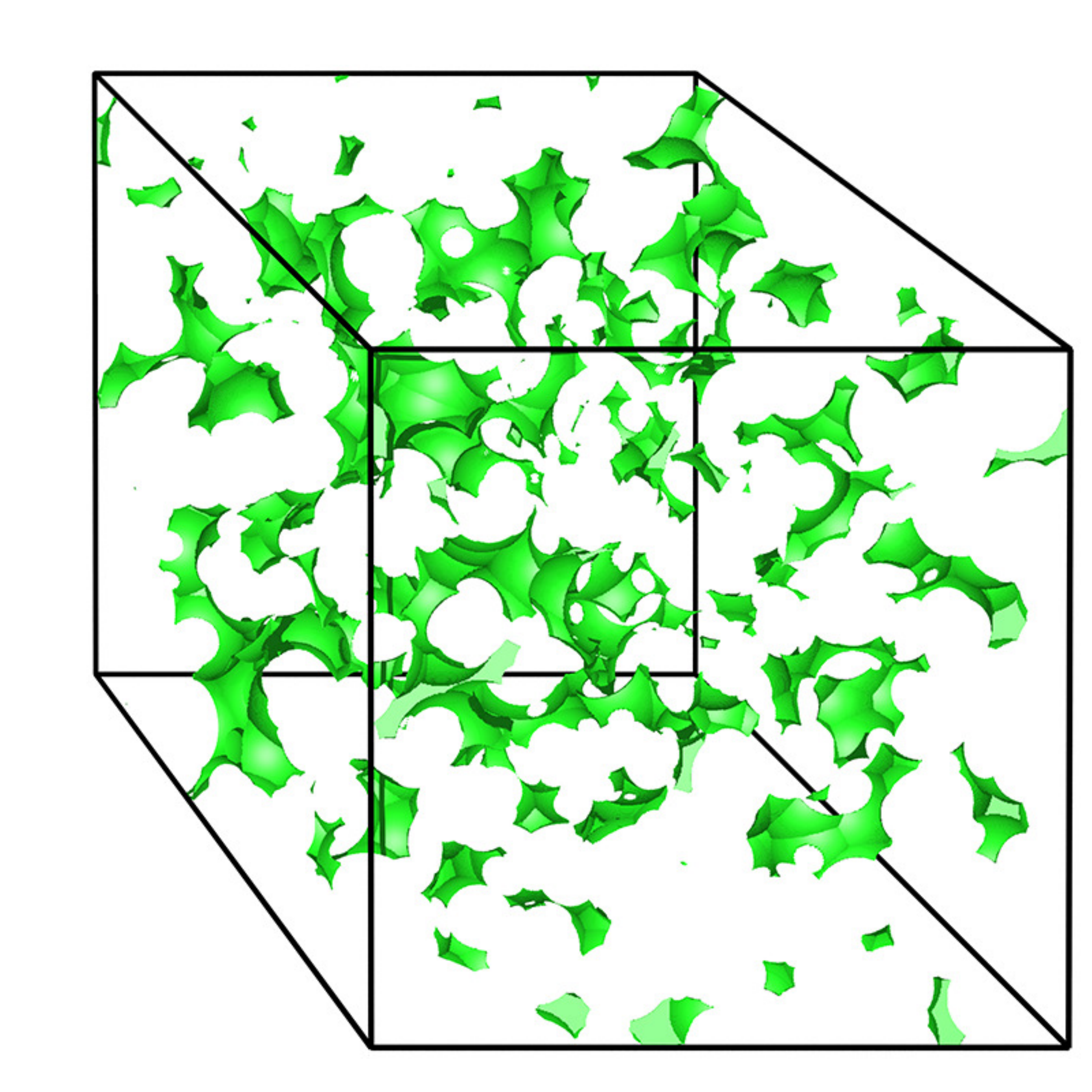}
\includegraphics[width=0.48\textwidth]{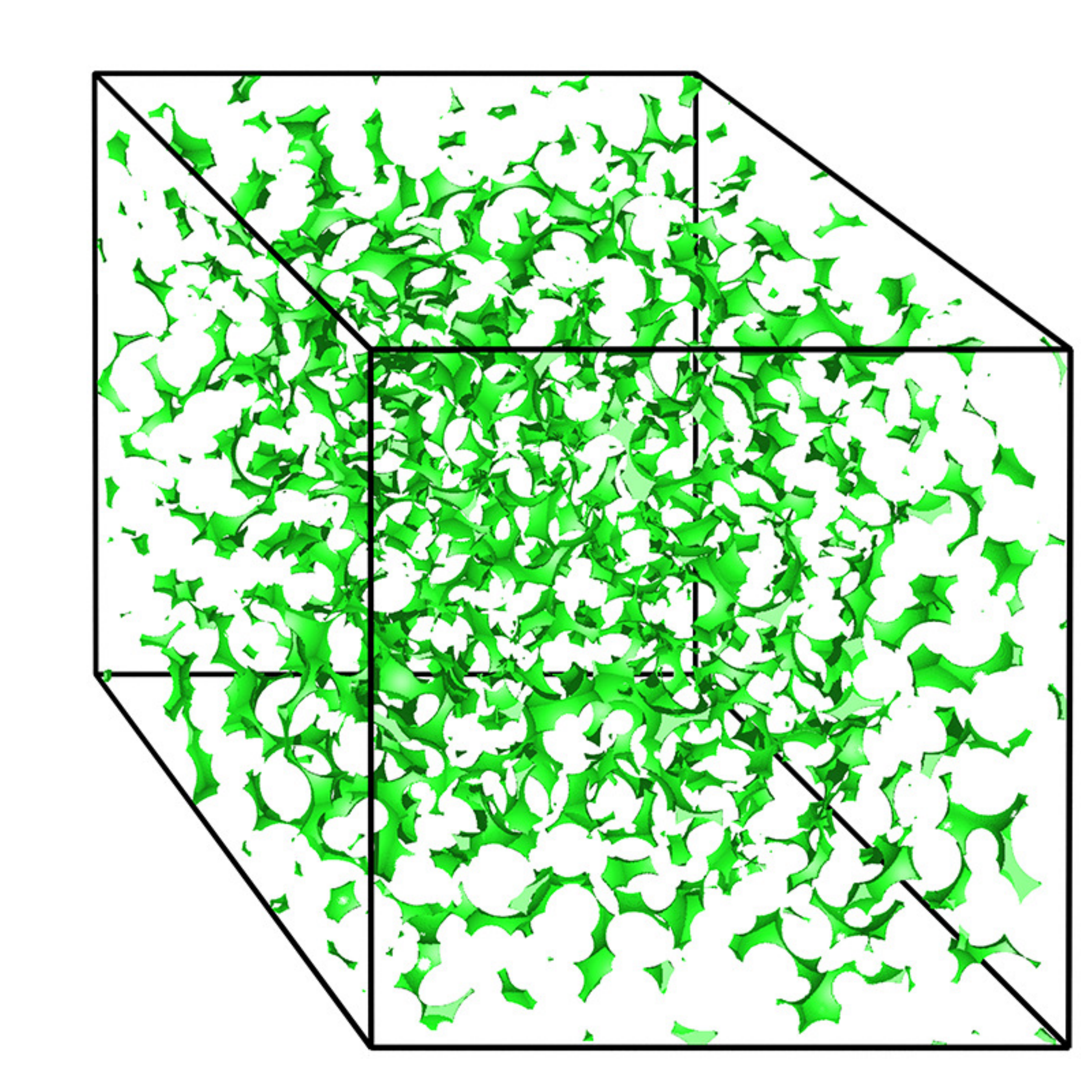}
\end{center}
\caption{Void phase in decorated stealthy ground states in three dimensions at $\chi=0.02$ (top) and $\chi=0.4598...$ (bottom), at the void-phase percolation threshold $a_{1c}=0.8970$ (top) or 0.6992 (bottom).}
\label{Configurations_32}
\end{figure}

We present visualizations of our two-phase systems derived from decorated stealthy ground states in Figs.~\ref{Configurations_2}~-~\ref{Configurations_32}. In three dimensions, we present separate figures for the particle phase and the void phase for clarity. In the rest of the section, we present the above-mentioned properties of our two-phase systems, and compare them with decorated Poisson point process and hard-sphere point process.


\subsection{Packing fraction and stealthiness}
\label{Packing_Stealthy}

\begin{figure}[H]
\begin{center}
\includegraphics[width=0.48\textwidth]{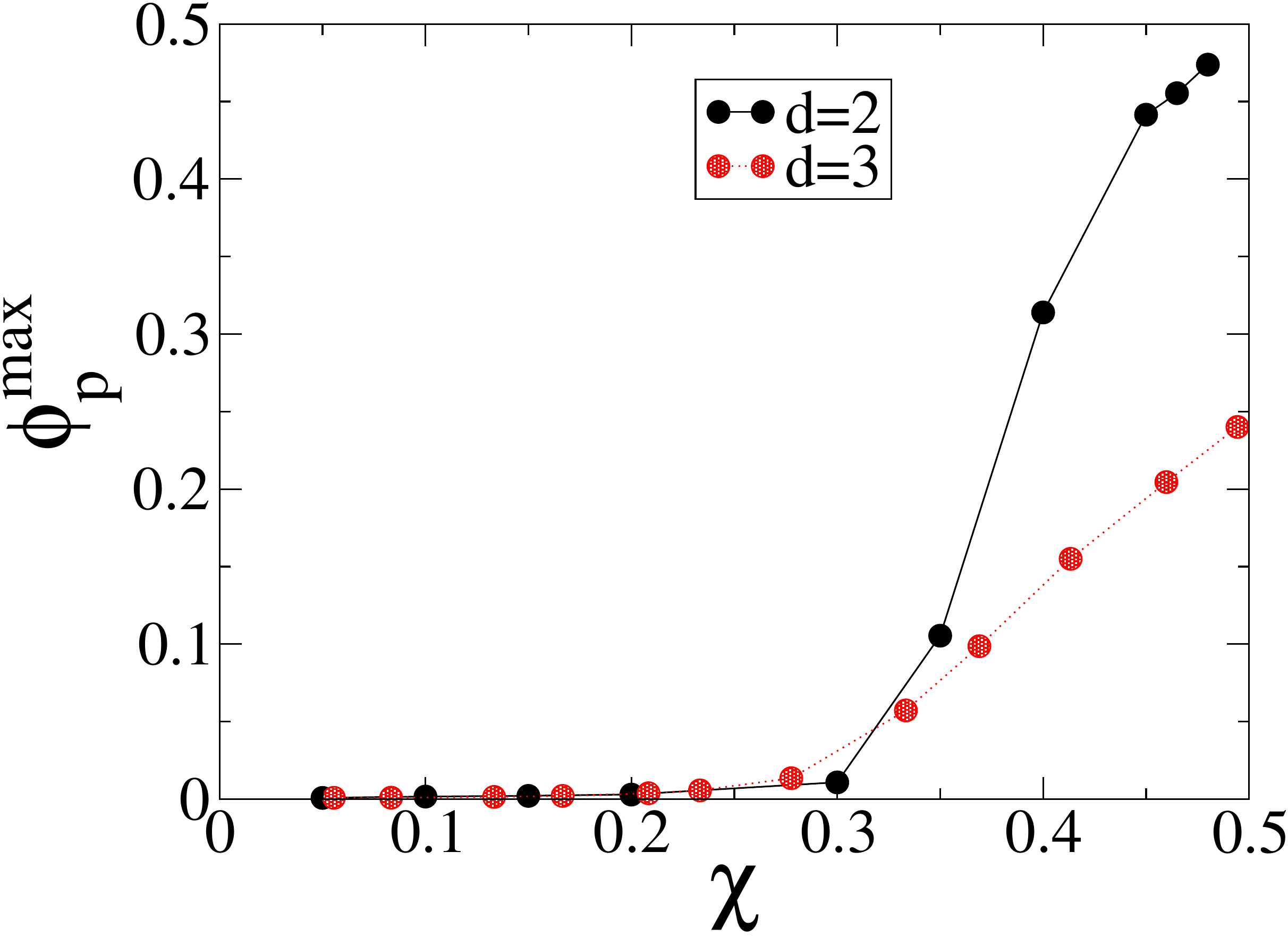}
\end{center}
\caption{Maximum packing fraction $\phi_p^{max}$, averaged over all configurations, of decorated stealthy ground states in two and three dimensions as a function of $\chi$.}
\label{PackingFraction}
\end{figure}

We present the maximum packing fraction of decorated stealthy ground states, $\phi_p^{max}$, in Fig.~\ref{PackingFraction}. In each dimension, as $\chi$ increases, $\phi_p^{max}$ remains to be zero for $\chi$ up to about 0.3 and then start to increase. This indicates that for $\chi \le 0.3$, particles in entropically favored stealthy ground states can become arbitrarily close to each other. As $\chi$ becomes higher, particles develop an effective hard core that are impenetrable. The development of such hard core was also observed in Ref.~\onlinecite{uche2004constraints}.

\begin{figure}[H]
\begin{center}
\includegraphics[width=0.48\textwidth]{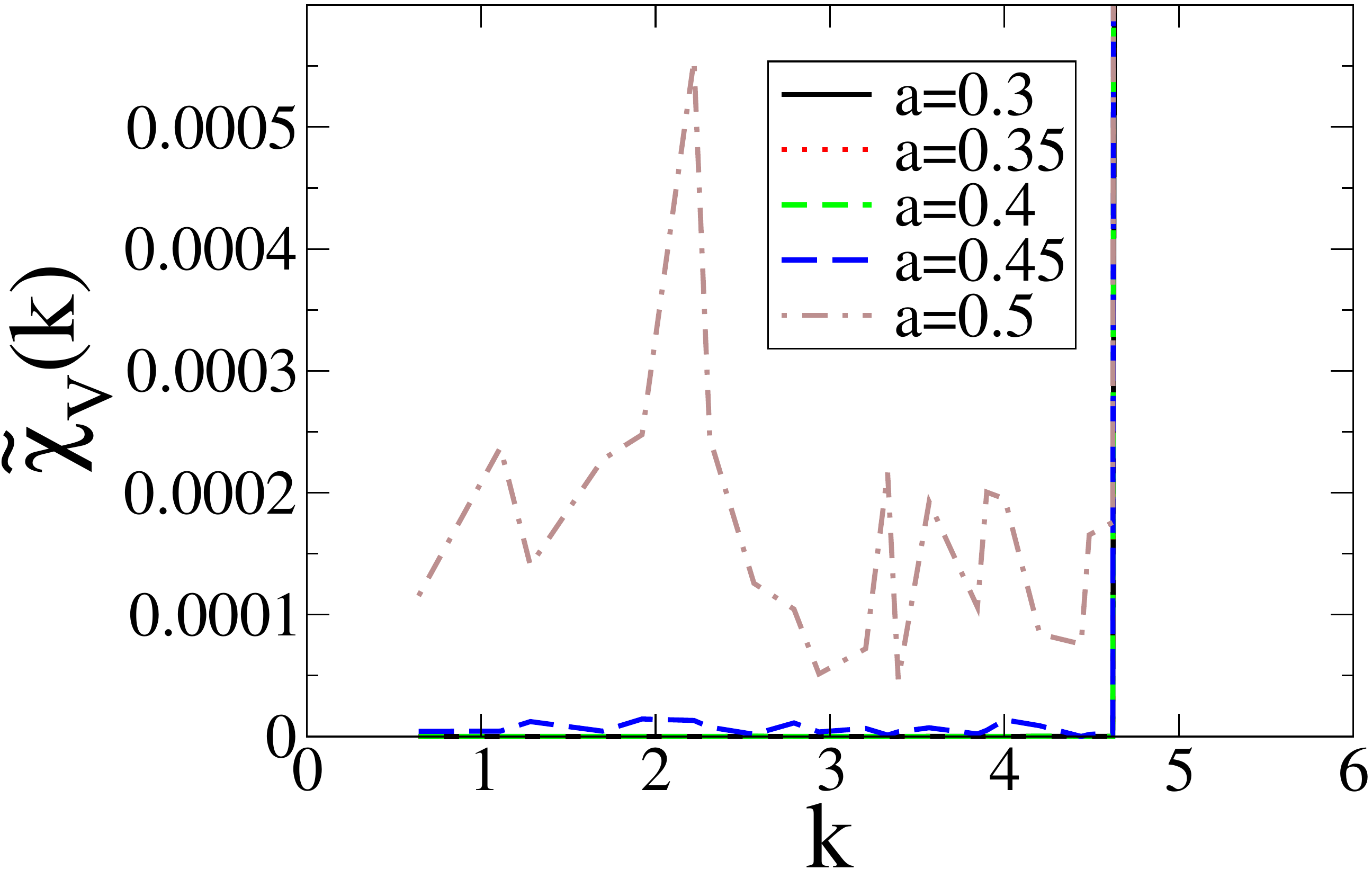}
\end{center}
\caption{Spectral density $\tilde \chi_{_V}(k)$ of a two-phase medium obtained from decorating a two-dimensional stealthy ground state with $N=111$ particles at $\chi=0.45$ with several different sphere radii $a$. }
\label{Stealthy_SpectralDensity}
\end{figure}

When we decorate a stealthy ground state and map it into a two-phase medium, if $\phi_2 \le \phi_p^{max}$, then Eq.~(\ref{chi_S}) ensures that the resulting two-phase medium is also stealthy. However, if $\phi_2 > \phi_p^{max}$, will the resulting two-phase medium  also be stealthy or hyperuniform? To answer this question, we decorated a two-dimensional stealthy ground state of $N=111$ particles at $\chi=0.45$ with several different sphere radii $a$, digitized the resulting two-phase medium into $10000\times10000$ pixels, and calculated the spectral density $\tilde \chi_{_V}(k)$ using Eq.~(\ref{chi_v}). The result is presented in Fig.~\ref{Stealthy_SpectralDensity}. For this particular system, the maximum packing radius is $a_p^{max}=0.407$. We see that for $a<a_p^{max}$, $\tilde \chi_{_V}(k)$ is zero for $k<4.7$. For $a>a_p^{max}$, however, $\tilde \chi_{_V}(k)$ is positive and does not tend to zero as $k\to 0$. Therefore, a decorated stealthy ground state is generally neither stealthy nor hyperuniform if $\phi_2 > \phi_p^{max}$.

\subsection{Effective diffusion coefficient}
\begin{figure}[H]
\begin{center}
\includegraphics[width=0.48\textwidth]{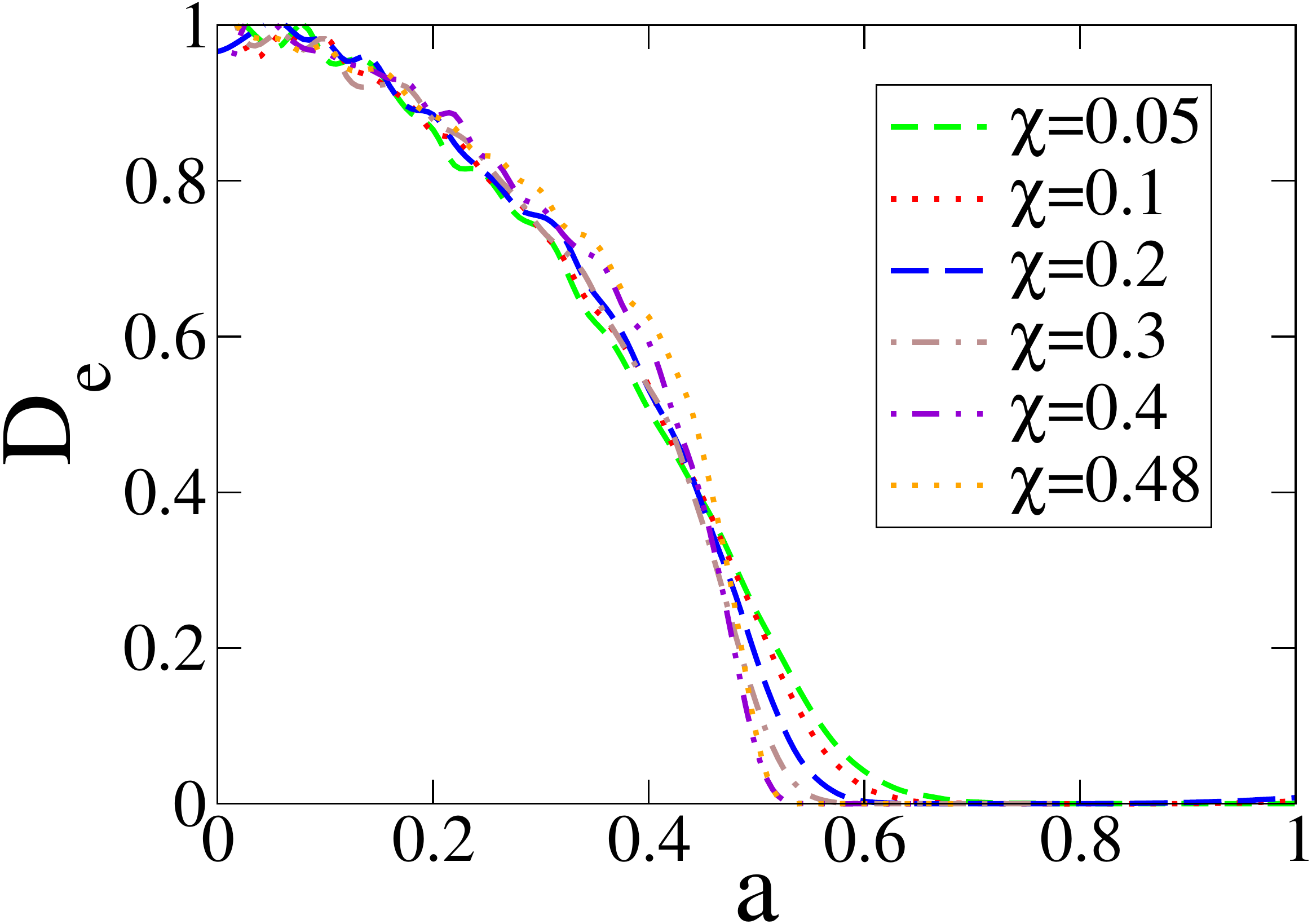}
\includegraphics[width=0.48\textwidth]{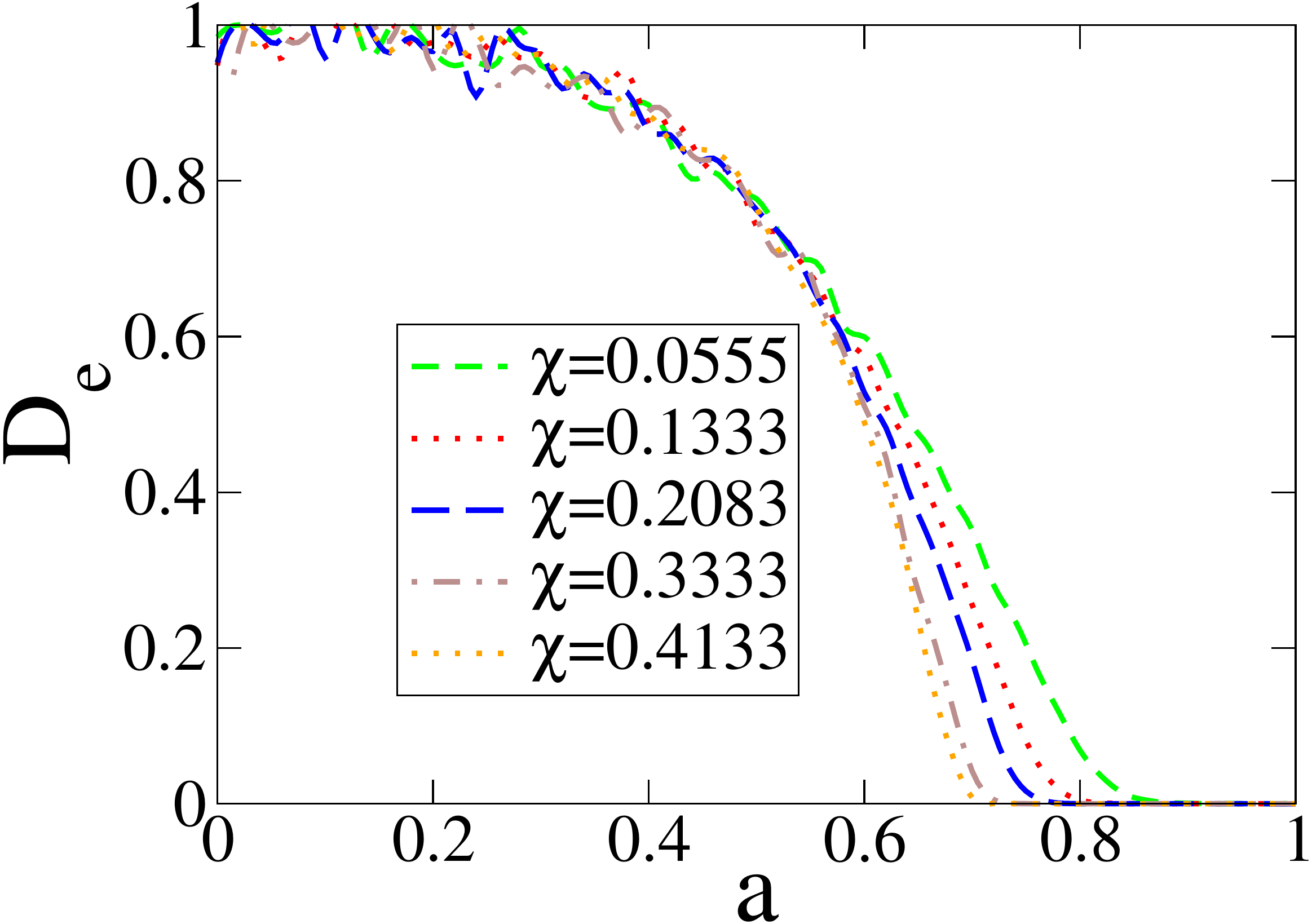}
\end{center}
\caption{The effective diffusion coefficient, $D_e$, for our two-phase systems derived from decorated stealthy ground states in two (top) and three (bottom) dimensions as a function of sphere radius $a$. The number density $\rho$ is fixed to be unity.}
\label{DiffusionCoefficient}
\end{figure}
We present the calculated effective diffusion coefficient for our two-phase systems derived from decorated stealthy ground states in Fig.~\ref{DiffusionCoefficient}. It is interesting to note that in two dimensions, the curves of $D_e(a)$ cross over each other for different values of $\chi$: while for smaller $a$ higher $\chi$ produces a higher $D_e$, for larger $a$ higher $\chi$ produces a smaller $D_e$. An explanation for such phenomenon will be presented in the next paragraph. 

\begin{figure}[H]
\begin{center}
\includegraphics[width=0.48\textwidth]{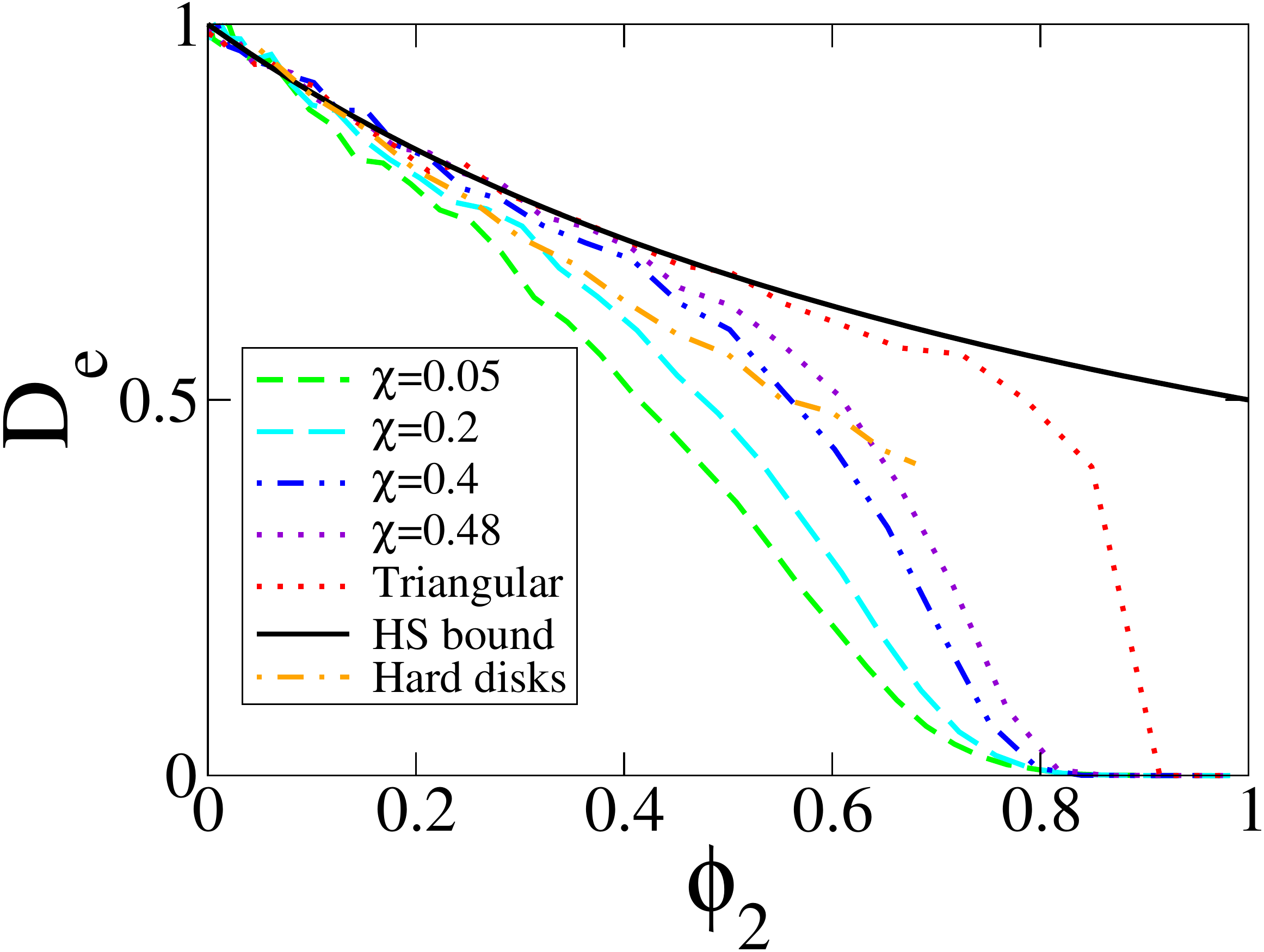}
\includegraphics[width=0.48\textwidth]{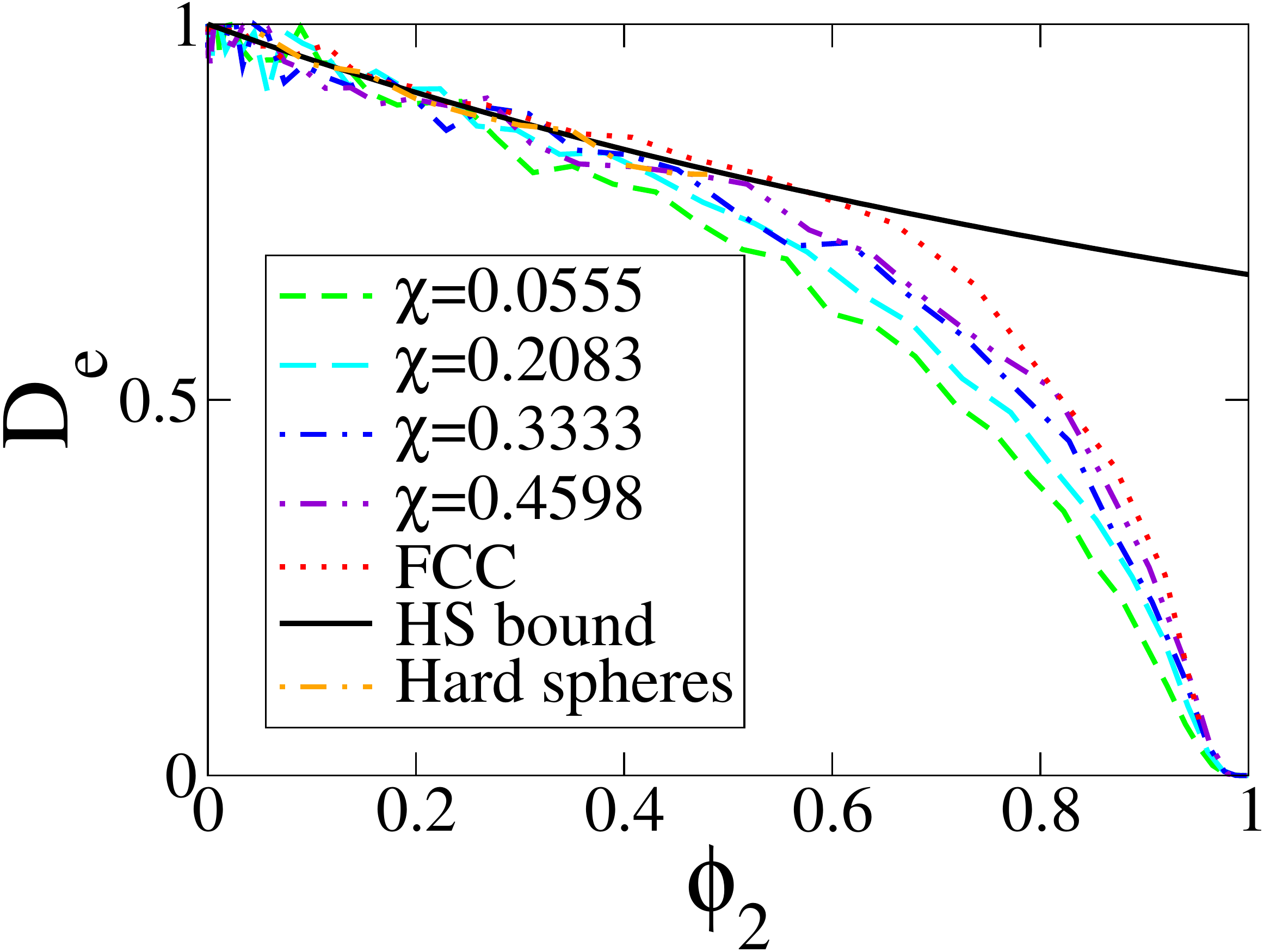}
\end{center}
\caption{The effective diffusion coefficient, $D_e$, for our two-phase systems derived from decorated stealthy ground states in two (top) and three (bottom) dimensions as a function of particle-phase volume fraction $\phi_2$. The number density $\rho$ is fixed to be unity. The optimal Hashin-Shtrikman (HS) upper bound, $D_e$ for triangular lattice and face-centered cubic (FCC) lattice, and $D_e$ for equilibrium hard disks and spheres are also plotted.}
\label{DiffusionCoefficient2}
\end{figure}

It is also useful to plot $D_e$ versus the particle-phase volume fraction, $\phi_2$, by mapping $a$ to $\phi_2$ using Eq.~(\ref{Ev_phi}). We present such plots in Fig.~\ref{DiffusionCoefficient2}. These plots show that higher $\chi$ values (more ordered arrangements of the obstacle phase) always produce higher $D_e$ at the same volume fraction, which is consistent with our intuition: a more ordered arrangement of the obstacles leaves more space between them, and produces a higher $D_e$. So why did we see the opposite relationship between $\chi$ and $D_e$ in Fig.~\ref{DiffusionCoefficient}, except for smaller $a$ in 2D? It turns out that a lower $\chi$ induces more overlap between the spherical obstacles and thus results in a lower $\phi_2$. This in turn produces a higher $D_e$.

With $D_e$ plotted versus  $\phi_2$, it is interesting to compare our result with the HS upper bound given in Eq.~(\ref{HSBound}).
We make such comparison in Fig.~\ref{DiffusionCoefficient2}. Our result is consistent with the upper bound for any $\chi$ and $\phi_2$ except for small fluctuations, but the bound is sharp only for smaller $\phi_2$.

If it is desired to find structures that maximizes $D_e$, then any one of the degenerate 
structures that achieves the HS bound is optimal.
We see that our two-phase systems derived from decorated stealthy ground states at very high $\chi$'s are very close to being optimal for $\phi_2$ up to 0.4-0.5. In two dimensions, this $\phi_2$ range coincides with $\phi_p^{max}$ at high $\chi$'s. Since decorated stealthy ground states loses stealthiness as $\phi_2$ increases beyond $\phi_p^{max}$, our results suggest that a loss of stealthiness causes $D_e$ to stop being optimal. In three dimensions, however, $D_e$ is less sensitive to structures. Thus, although decorated stealthy ground states (at high $\chi$'s) stops being a packing at around $\phi_2=0.2$, $D_e$ does not deviate from the optimal value until about $\phi_2=0.5$. Our observation that $D_e$ is less sensitive to structures in 3D than in 2D is consistent with the trends indicated in Ref.~\onlinecite{torquato1998effective}, which found that in the infinite-$d$ limit, $D_e$ is given exactly by the arithmetic average of the diffusion coefficients of the two phases (weighted by their volume fraction), independent of the structure.

In Fig.~\ref{DiffusionCoefficient2} we also present $D_e$ of decorated lattice structures (i.e., periodic arrays of spherical inclusions). Since these lattice structures are stealthy with even higher $\chi$ values, unsurprisingly, their $D_e$ sticks with the HS upper bound for an even larger $\phi_2$ range.

Lastly, we would like to mention a difference between the support of $D_e$ as a function of $\phi_2$ in 2D versus 3D. While $D_e$ for for our two-phase systems in 2D diminishes to zero at $\phi_2 \approx 0.8$, in 3D $D_e$ does not vanish until $\phi_2 \approx 0.97$. This difference emerges from the difference in the topological (connectedness) characteristics of the void phase between these dimensions. 

\subsection{Survival probability and mean survival time}

\begin{figure}[H]
\begin{center}
\includegraphics[width=0.4\textwidth]{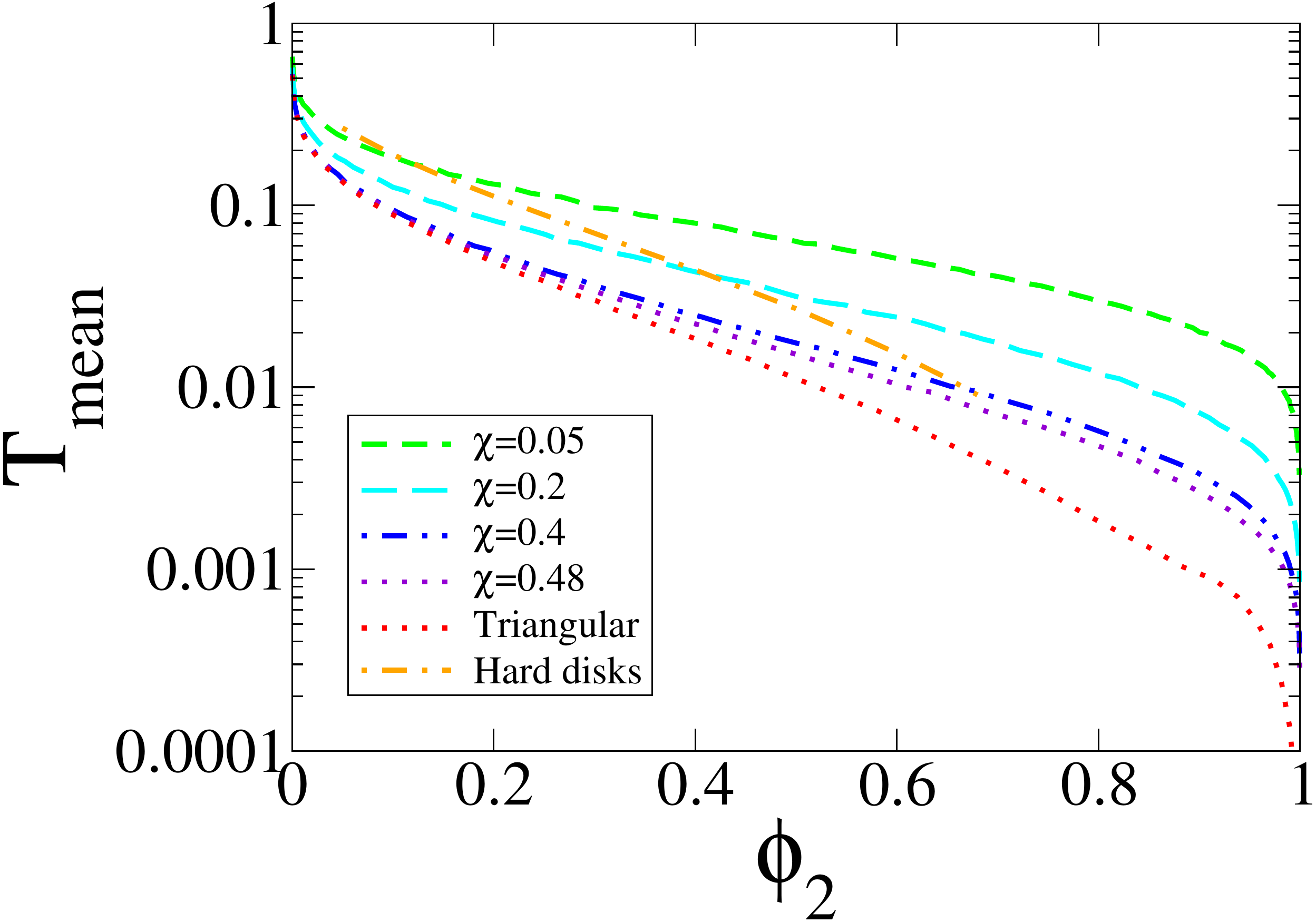}
\includegraphics[width=0.4\textwidth]{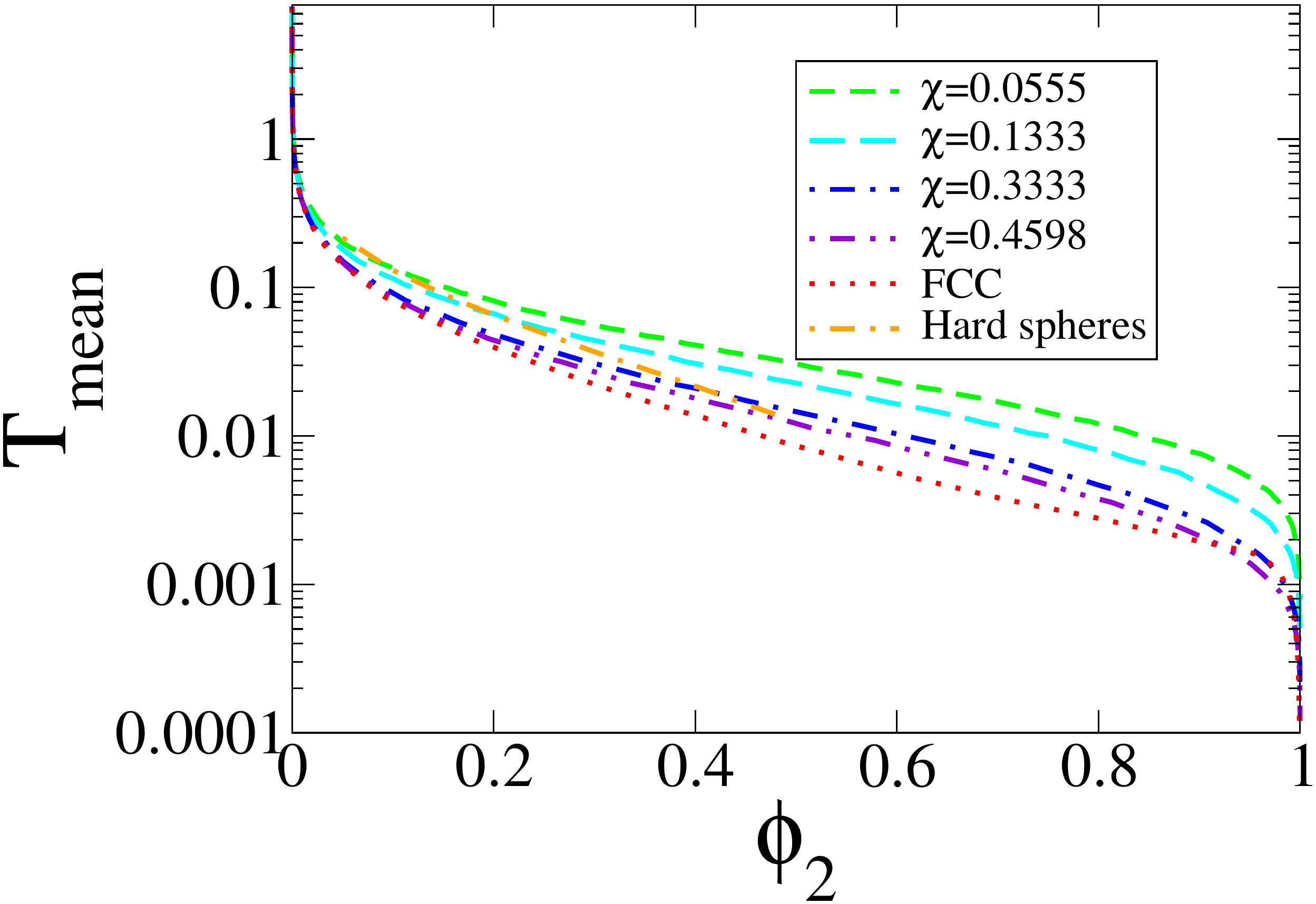}
\end{center}
\caption{The mean survival time, $T_{mean}$, as a function of particle-phase volume fraction $\phi_2$, for our two-phase systems derived from decorated stealthy ground states in 2D (top) and 3D (bottom). The same quantity for equilibrium disordered hard-sphere system is also included for comparison. The number density $\rho$ is fixed to be unity.}
\label{MeanSurvivalTime}
\end{figure}

\begin{figure}[H]
\begin{center}
\includegraphics[width=0.4\textwidth]{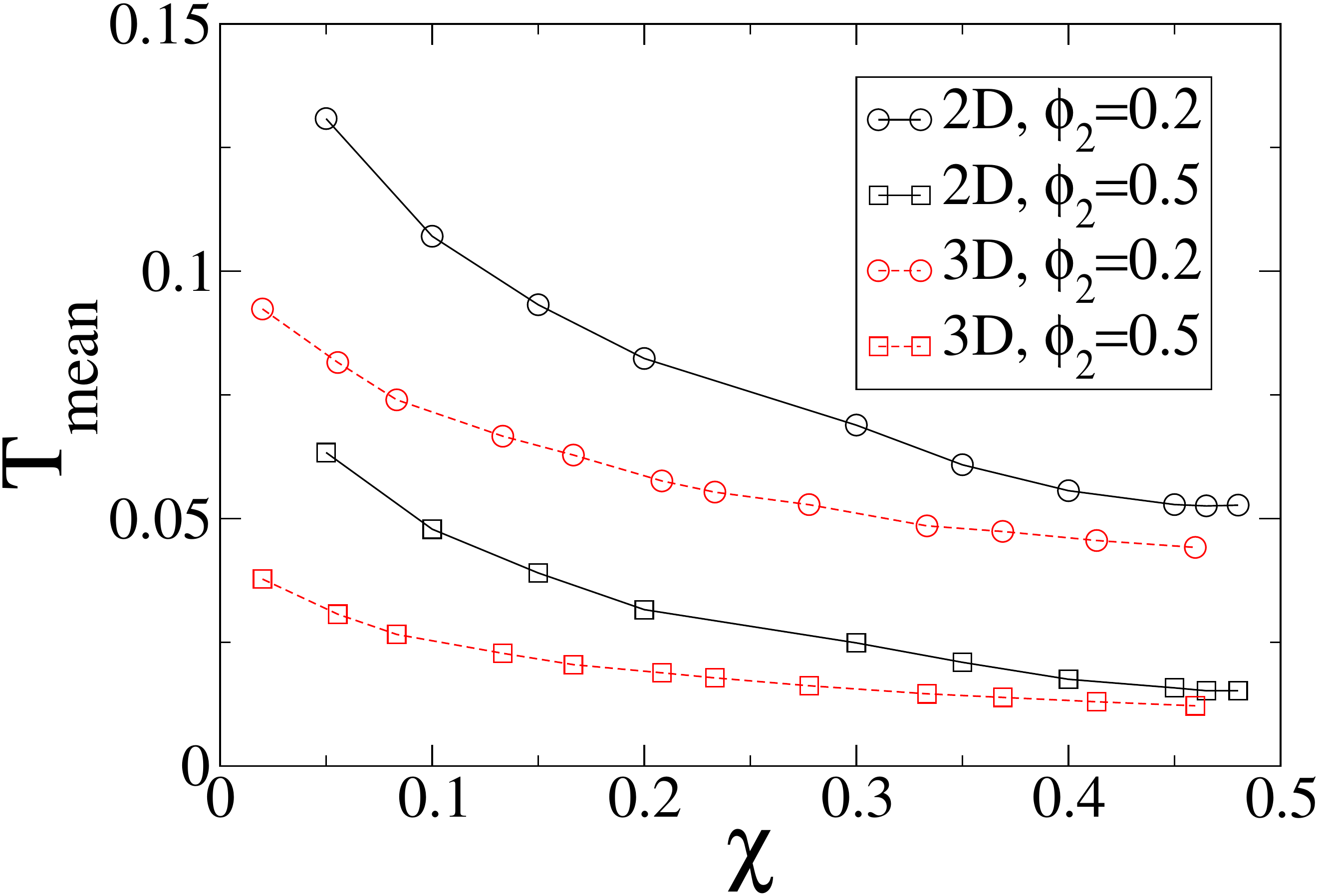}
\end{center}
\caption{The mean survival time $T_{mean}$ for our two-phase systems derived from decorated stealthy ground states in two and three dimensions at phase 2 volume fraction $\phi_2=0.2$ and 0.5. The number density $\rho$ is fixed to be unity.}
\label{MeanSurvivalTime2}
\end{figure}

We have computed the mean survival time, $T_{mean}$, of a diffusing reactant with unit diffusion coefficient, in our two-phase systems derived from decorated stealthy ground states. These results are summarized in Fig.~\ref{MeanSurvivalTime}. For comparison, the same quantity for equilibrium disordered hard-sphere systems are also included. Clearly, increasing order (increasing $\chi$ for stealthy ground states or increasing $\phi_2$ for hard spheres) suppresses $T_{mean}$. However, there is a crossover between the curves for stealthy ground states and that for equilibrium disordered hard spheres. 
This crossover is expected because as $\phi_2$ increases, an equilibrium hard-sphere system becomes more ordered, and therefore comparable to a stealthy two-phase medium with a higher $\chi$.
In Fig.~\ref{MeanSurvivalTime2}, we plot $T_{mean}$ versus $\chi$ for $\phi_2=0.2$ and 0.5. We see that in 2D, $T_{mean}$ is somewhat more sensitive to $\chi$ than in 3D.



\begin{figure}[H]
\begin{center}
\includegraphics[width=0.48\textwidth]{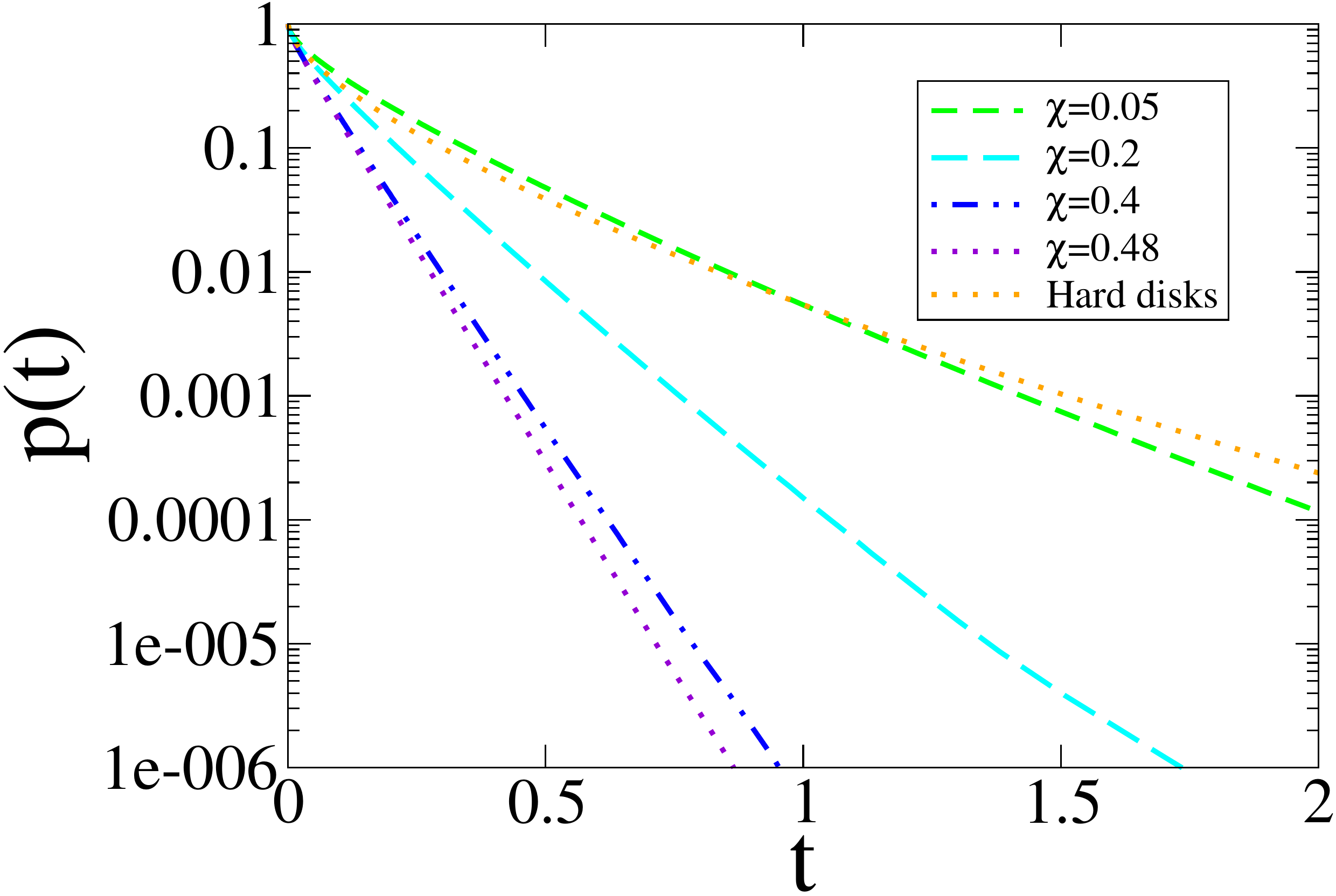}
\includegraphics[width=0.48\textwidth]{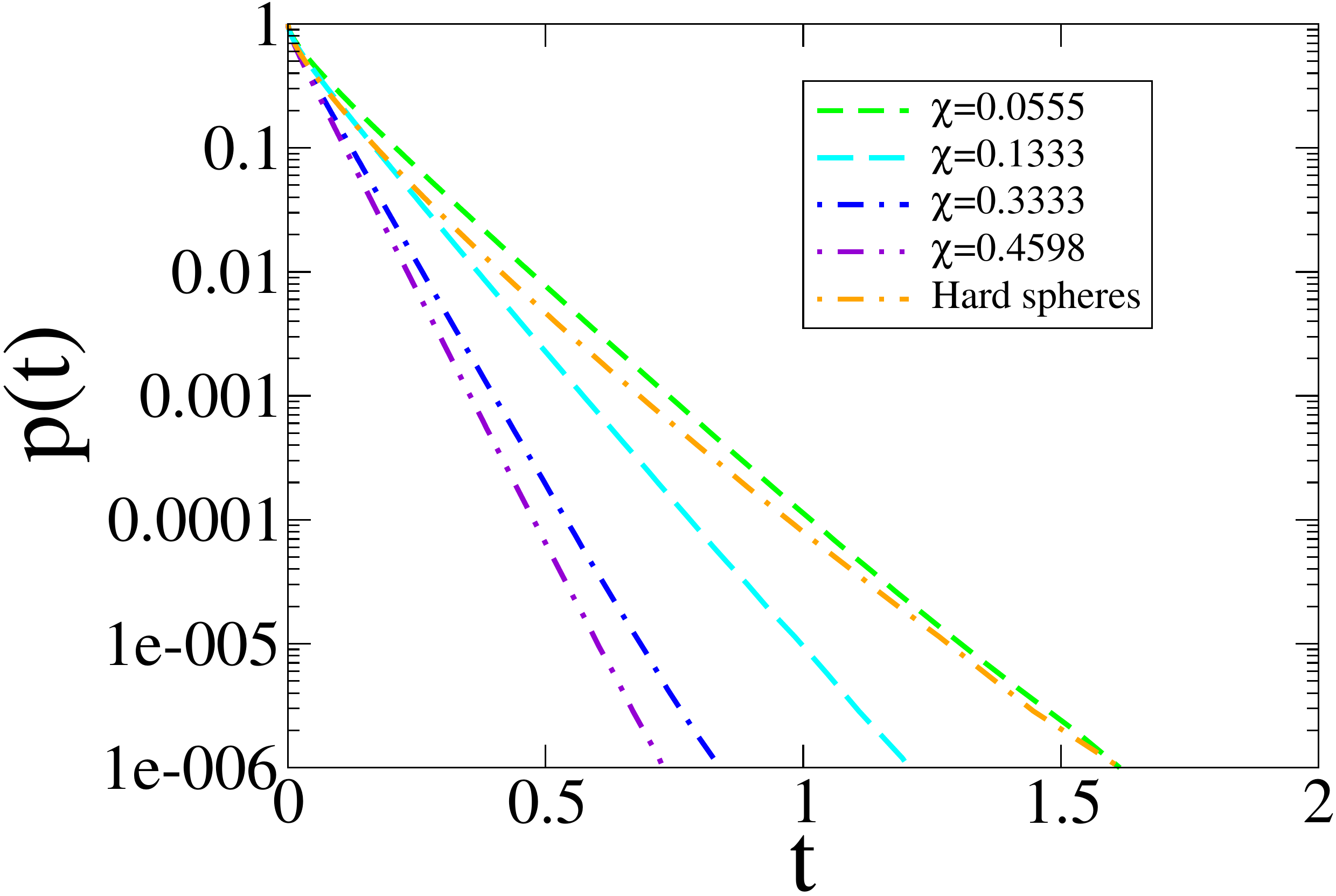}
\end{center}
\caption{The survival probability $p(t)$ for our two-phase systems derived from decorated stealthy ground states in 2D (top) and 3D (bottom) at phase 2 volume fraction $\phi_2=0.2$. The same quantity for equilibrium disordered hard-sphere system is also included for comparison. The number density $\rho$ is fixed to be unity.}
\label{SurvivalProbability}
\end{figure}

In Fig.~\ref{SurvivalProbability} we present the survival probability, $p(t)$, at $\phi_1=0.5$, for our two-phase systems derived from decorated stealthy ground states and equilibrium disordered hard spheres. The same crossover phenomenon also appears here, suggesting that the long-range order possessed by stealthy ground states suppresses $p(t)$ at large $t$ more efficiently, while the short-range order possessed by equilibrium disordered hard spheres suppresses $p(t)$ at small $t$ more efficiently.

\begin{figure}[H]
\begin{center}
\includegraphics[width=0.48\textwidth]{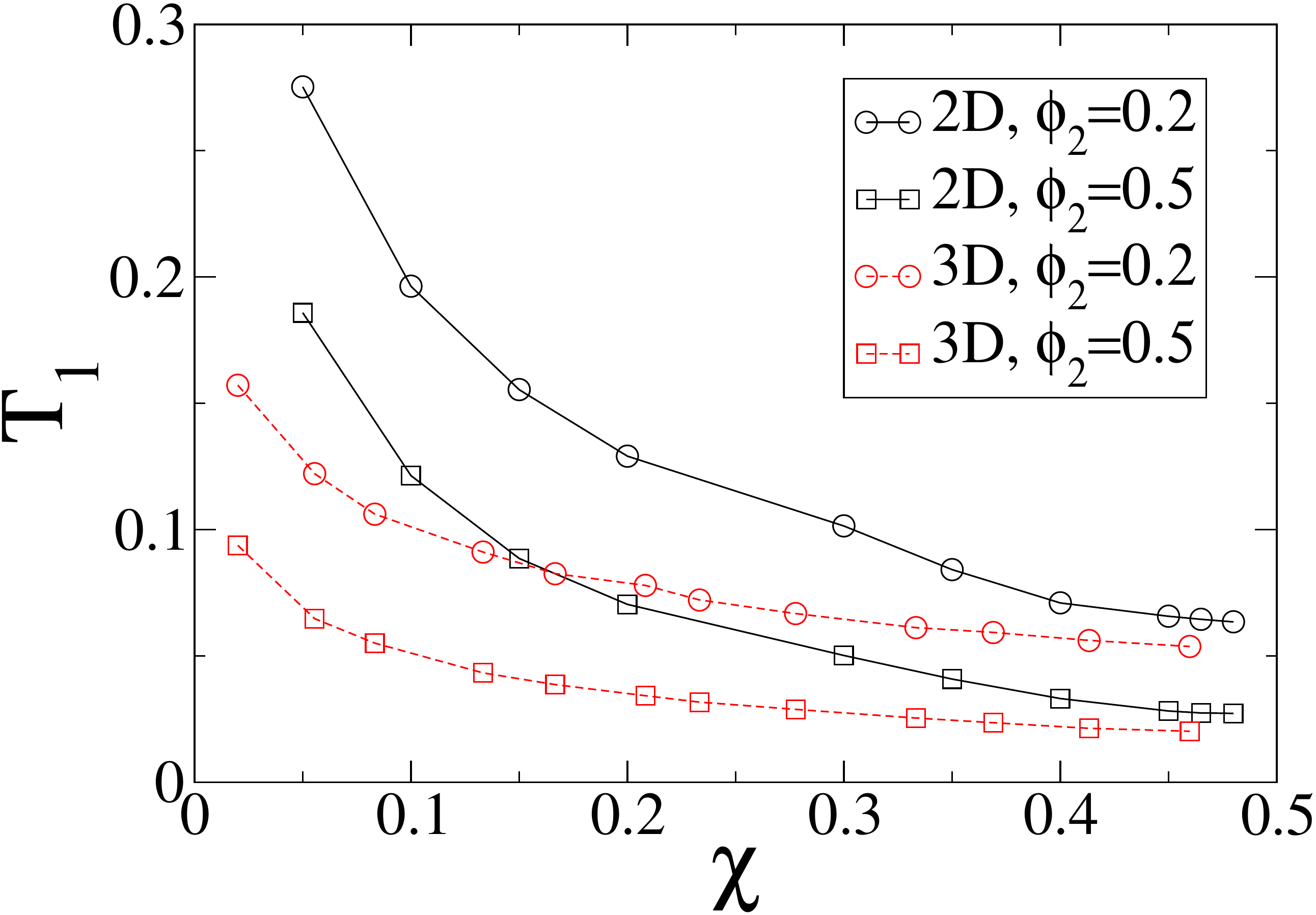}
\end{center}
\caption{Principal relaxation time $T_1$ for our two-phase systems derived from decorated stealthy ground states in two and three dimensions at phase 2 volume fraction $\phi_2=0.2$ and 0.5. The number density $\rho$ is fixed to be unity.}
\label{RelaxationTime}
\end{figure}

\begin{figure}[H]
\begin{center}
\includegraphics[width=0.48\textwidth]{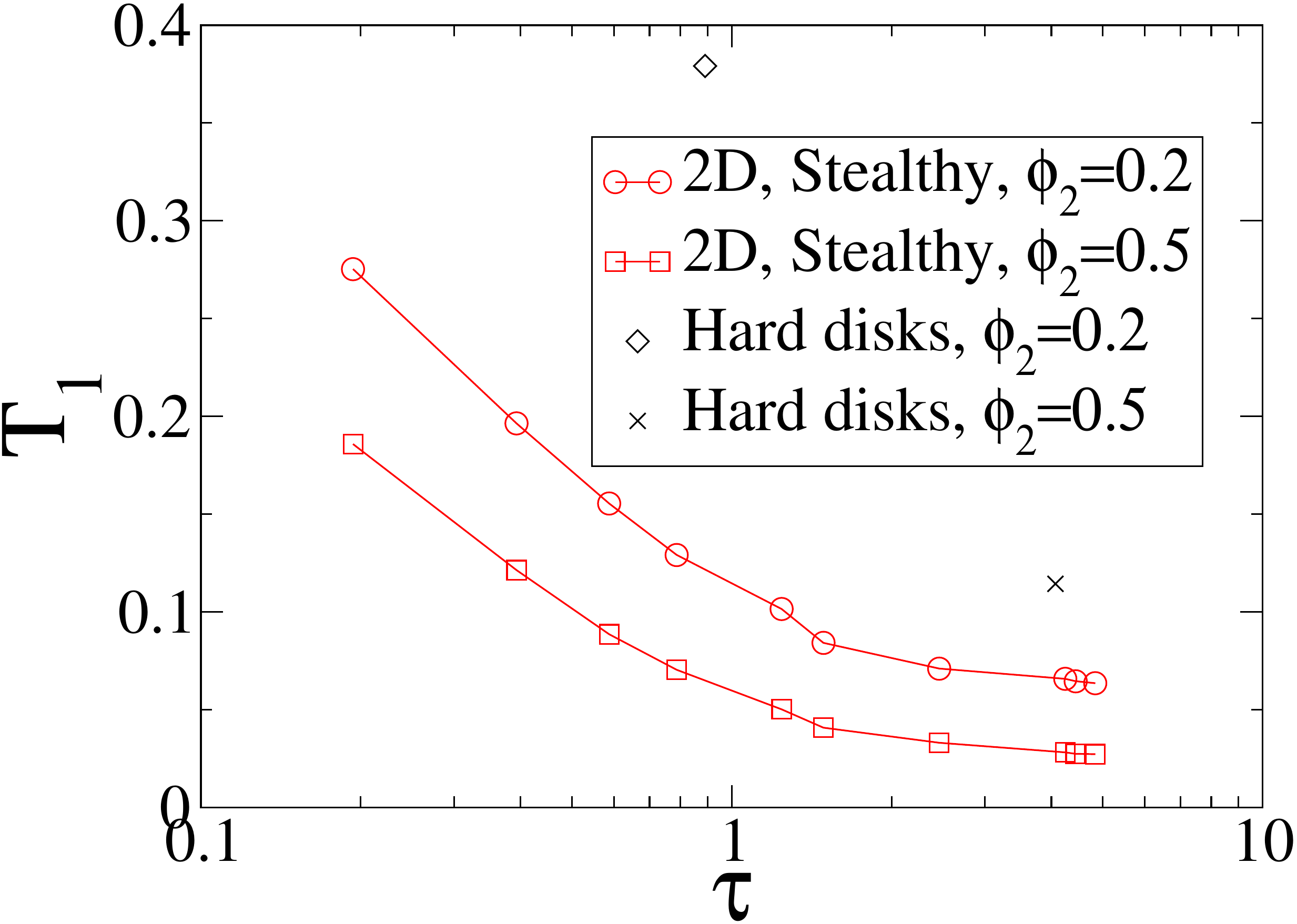}
\includegraphics[width=0.48\textwidth]{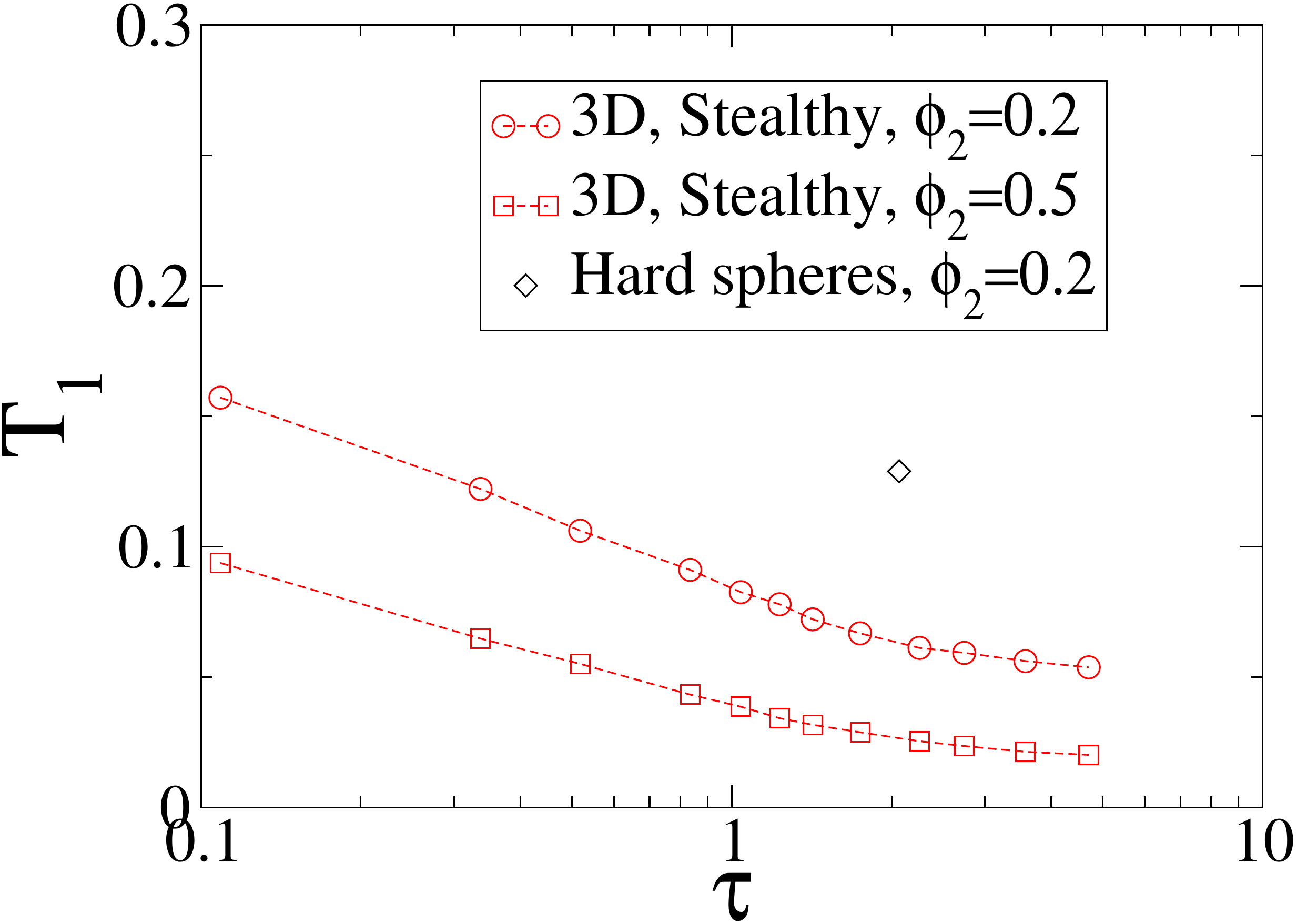}
\end{center}
\caption{Principal relaxation time $T_1$ for our two-phase systems derived from decorated stealthy ground states and equilibrium disordered hard spheres in 2D (top) and 3D (bottom) at volume fraction $\phi_2=0.2$ and 0.5. The number density $\rho$ is fixed to be unity.}
\label{RelaxationTime_tau}
\end{figure}

In Fig.~\ref{RelaxationTime} we present the principal relaxation time $T_1$ for our two-phase systems derived from decorated stealthy ground states. It turns out that $T_1$ is much more sensitive to $\chi$ in 2D than in 3D. More interestingly, one can compare $T_1$ of stealthy ground states and equilibrium disordered hard disks at the same order metric $\tau$. We present such comparison in Fig.~\ref{RelaxationTime_tau}. In two dimensions, one can see that at $\phi_2=0.2$, $T_1$ of equilibrium disordered hard disks is much higher than that of our two-phase systems derived from decorated stealthy ground states with similar $\tau$'s. As we explained earlier, $T_1$ is related to the pore-size distribution. Therefore, our results suggest that hyperuniformity suppresses the formation of large holes, even in the very disordered regime. As $\phi_2$ increases to 0.5, however, the difference between the two systems diminishes. Our finite-sized simulation results suggest that at this value of $\phi_2$, even equilibrium hard-sphere systems suppress the formation of large holes very well. \footnote{We should clarify that in the infinite-system-size limit, $T_1$ of equilibrium disordered hard-sphere systems is actually infinite because of a non-zero probability of forming arbitrarily large holes (i.e., $P(\delta)$ is non-zero for arbitrarily large $\delta$). \cite{torquato1991diffusion}
For finite-sized systems, however, $T_1$ is finite.
For decorated stealthy ground states, it is unclear whether or not $T_1$ would be infinite in the infinite-system-size limit.
} 

\subsection{Geometrical and Topological Properties}
\begin{figure}[H]
\includegraphics[width=0.5\textwidth]{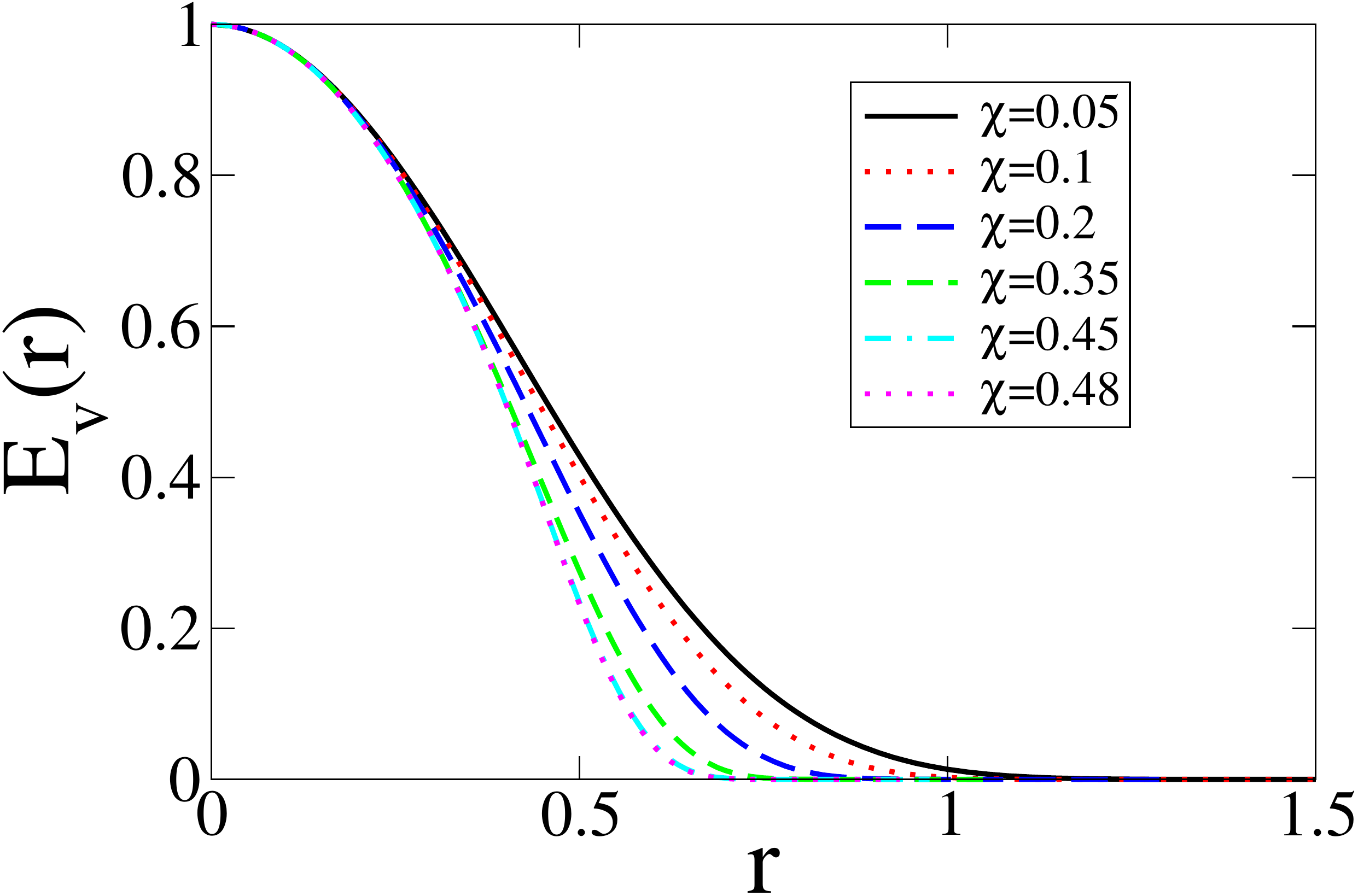}
\includegraphics[width=0.5\textwidth]{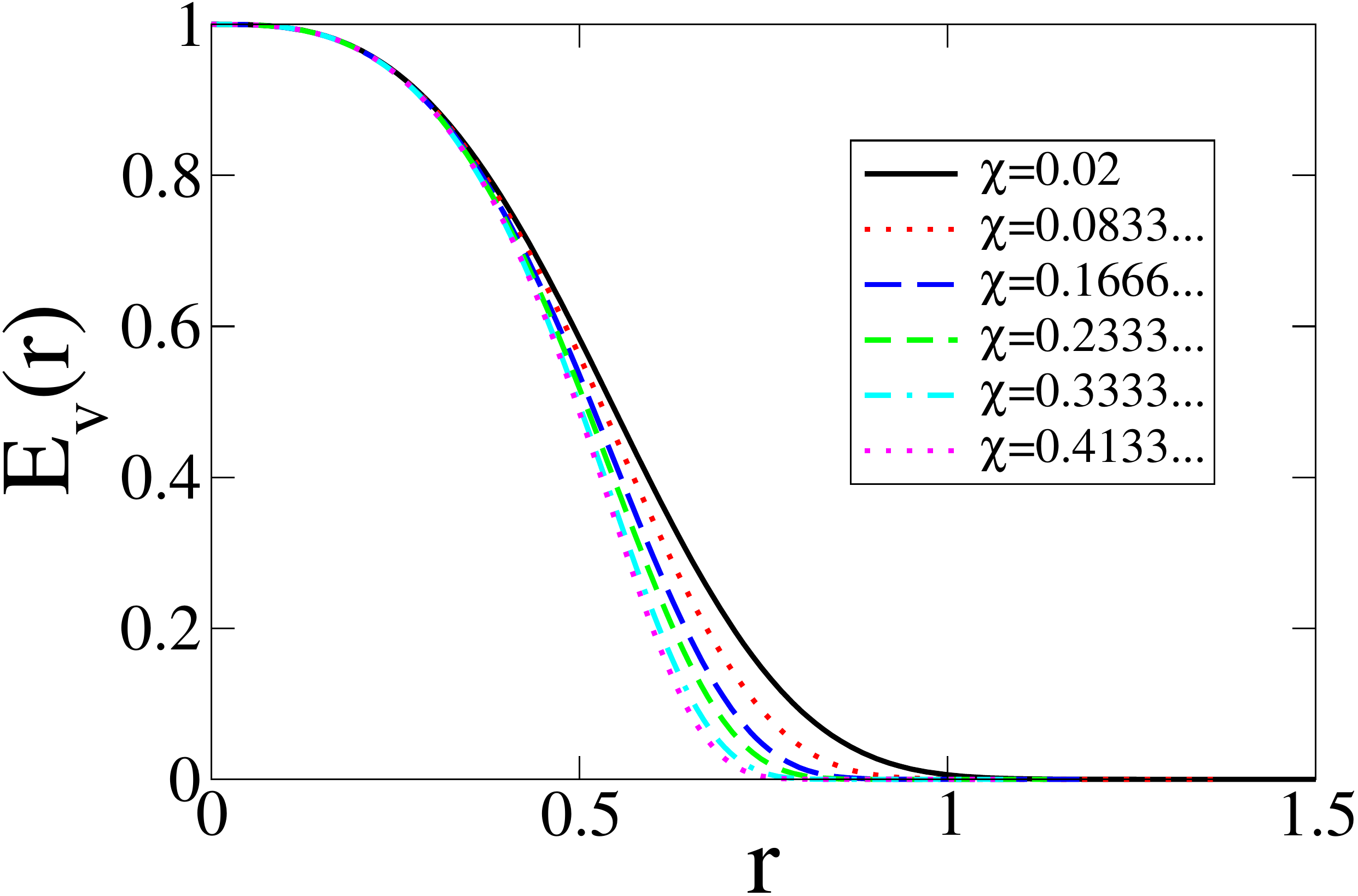}
\caption{Void-exclusion probability $E_V(r)$ of entropically favored stealthy ground states in two (top) and three (bottom) dimensions at unit number density.}
\label{Ev}
\end{figure}

\begin{figure}[H]
\begin{center}
\includegraphics[width=0.5\textwidth]{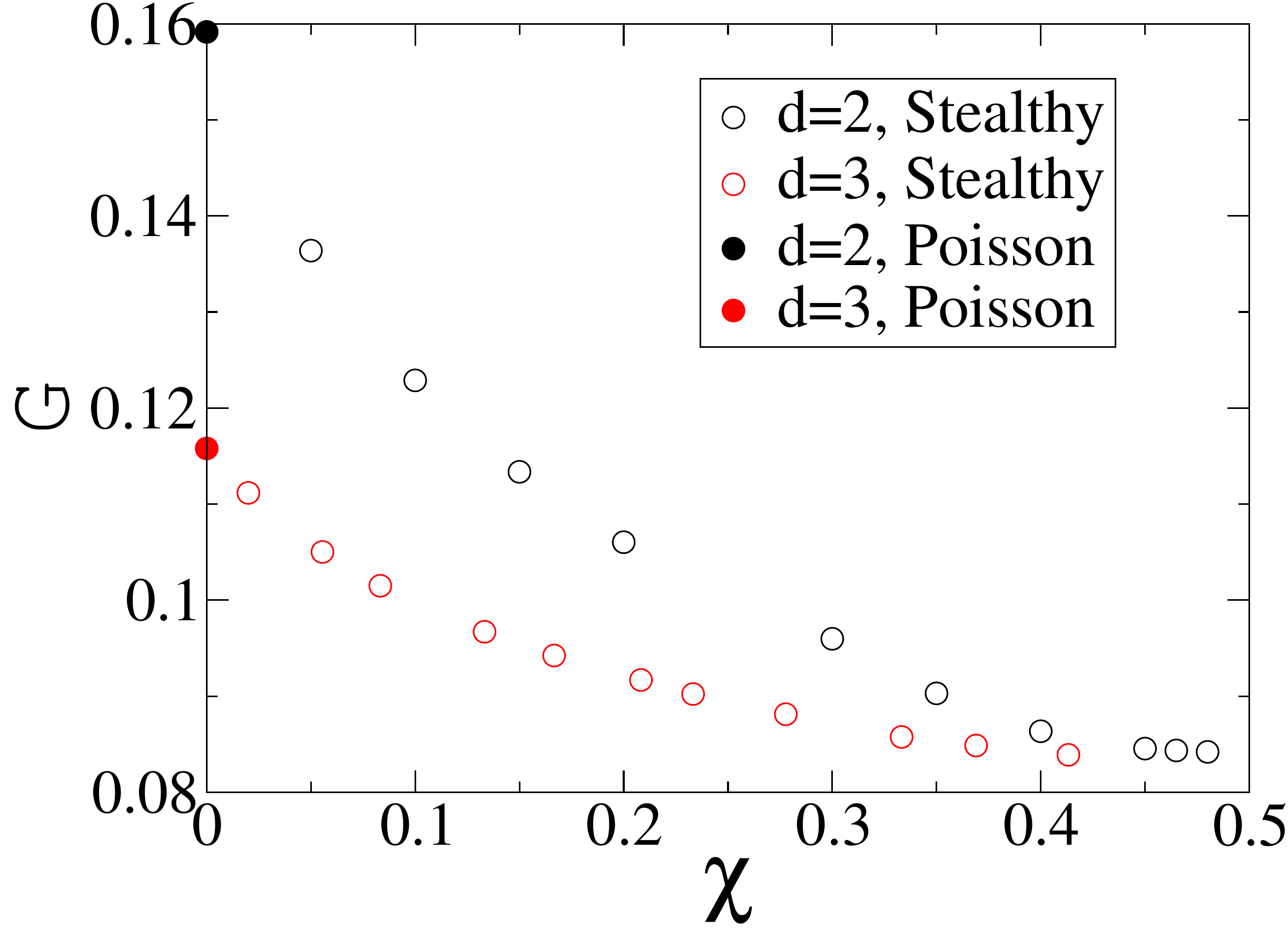}
\end{center}
\caption{Quantizer error $\mathcal{G}$ of entropically favored stealthy ground states in two and three dimensions at unit number density.}
\label{quantizererror}
\end{figure}

\begin{figure}[H]
\begin{center}
\includegraphics[width=0.5\textwidth]{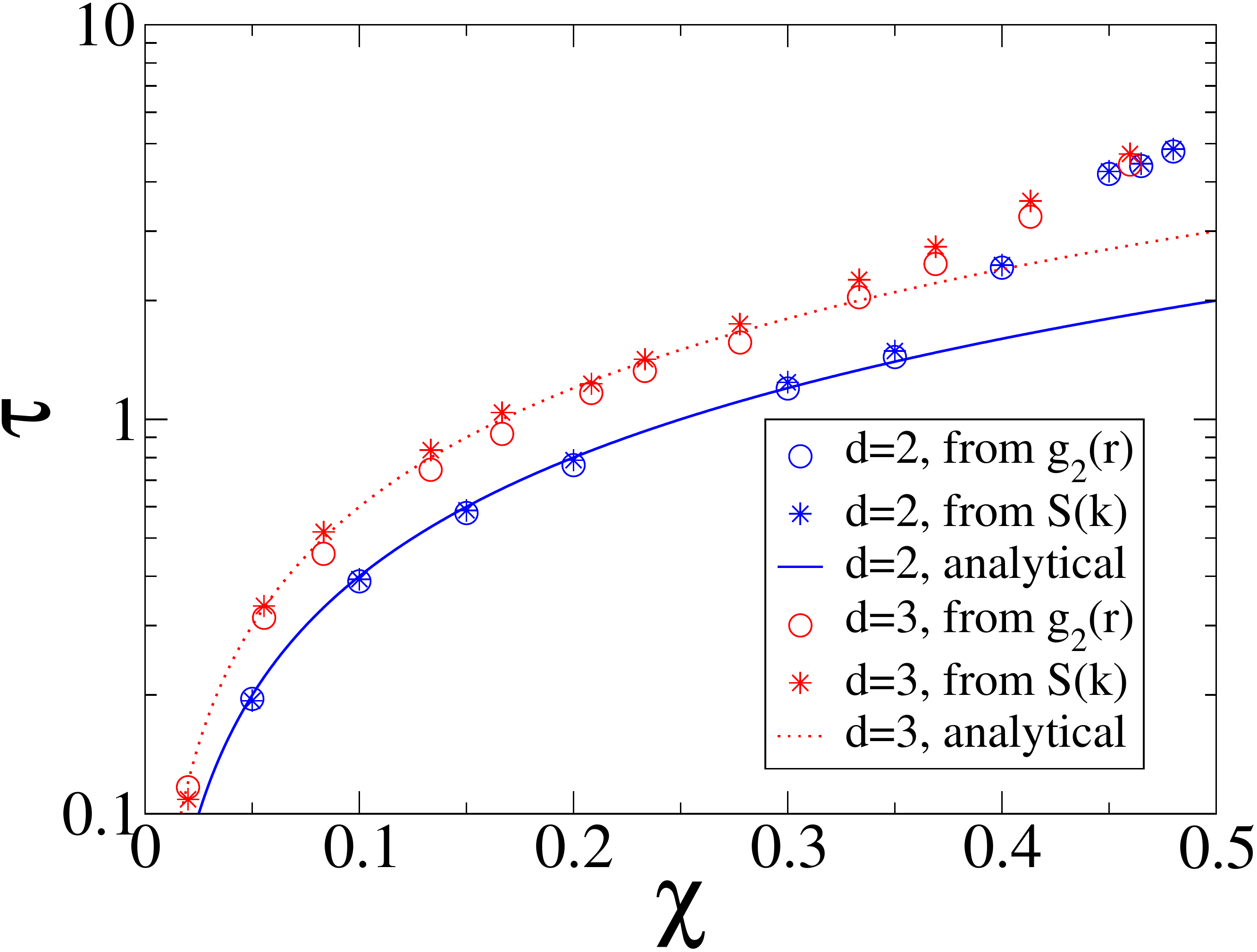}
\end{center}
\caption{Order metric $\tau$ of entropically favored stealthy ground states in two and three dimensions at unit number density, calculated from pair correlation function $g_2(r)$ and structure factor $S(k)$. We also include an analytical approximation for $\tau$, given in Ref. 18, which is $\tau=2 d \chi$.}
\label{ordermetric}
\end{figure}

The percolation volume fraction for both phases in 2D and 3D was already presented in Fig.~\ref{perco}~and~\ref{perco1}. The void-exclusion probability $E_V(r)$, quantizer error $\mathcal{G}$, and order metric $\tau$ are presented in Figs.~\ref{Ev}-\ref{ordermetric}. We see that in each dimension, as $\chi$ increases, $\phi_{c}$ increases, $E_V(r)$ at any $r$ decreases, $\mathcal{G}$ decreases, and $\tau$ increases.

The order metric $\tau$ can be computed from either $g_2(r)$ or $S(k)$, as shown in Fig.~\ref{ordermetric}. In 2D, the results from these two approaches have good consistency. However, in 3D, $\tau$ computed from $g_2(r)$ is often slightly lower than $\tau$ computed from $S(k)$. We discovered that this is because $g_2(r)$ is still oscillating around 1 at half the simulation box side length, where the integration in Eq.~(\ref{tau}) has to be cut off. Therefore, such a cutoff should make $\tau$ computed from $g_2(r)$ too low. We thus use $\tau$ computed from $S(k)$ in the rest of the paper. It is
seen that $\tau$ is very sensitive
at detecting the rise in short-range and long-range order as $\chi$ increases.

\subsection{Correlations between geometrical properties, and comparison with equilibrium disordered hard-sphere systems}

\begin{figure}[H]
\begin{center}
\includegraphics[width=0.48\textwidth]{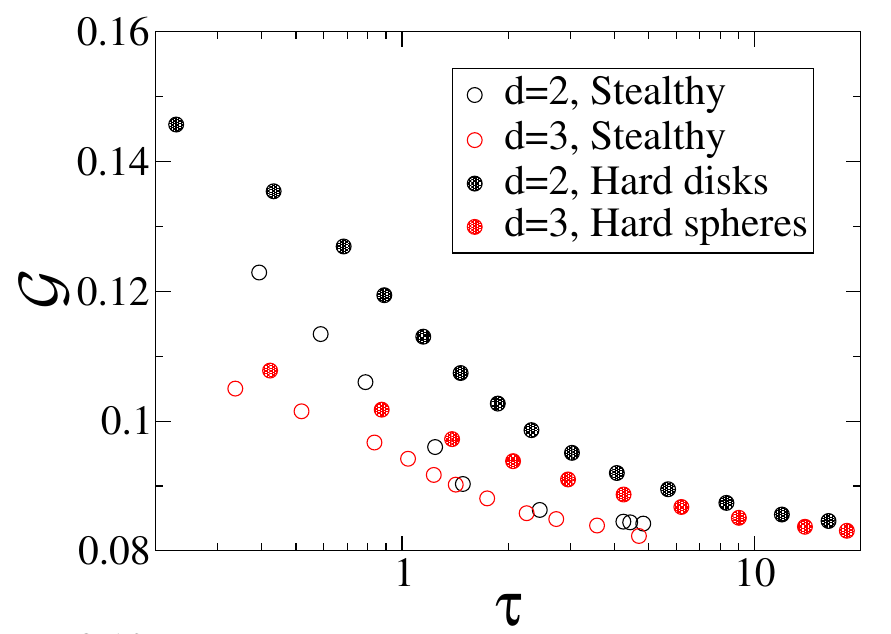}
\includegraphics[width=0.48\textwidth]{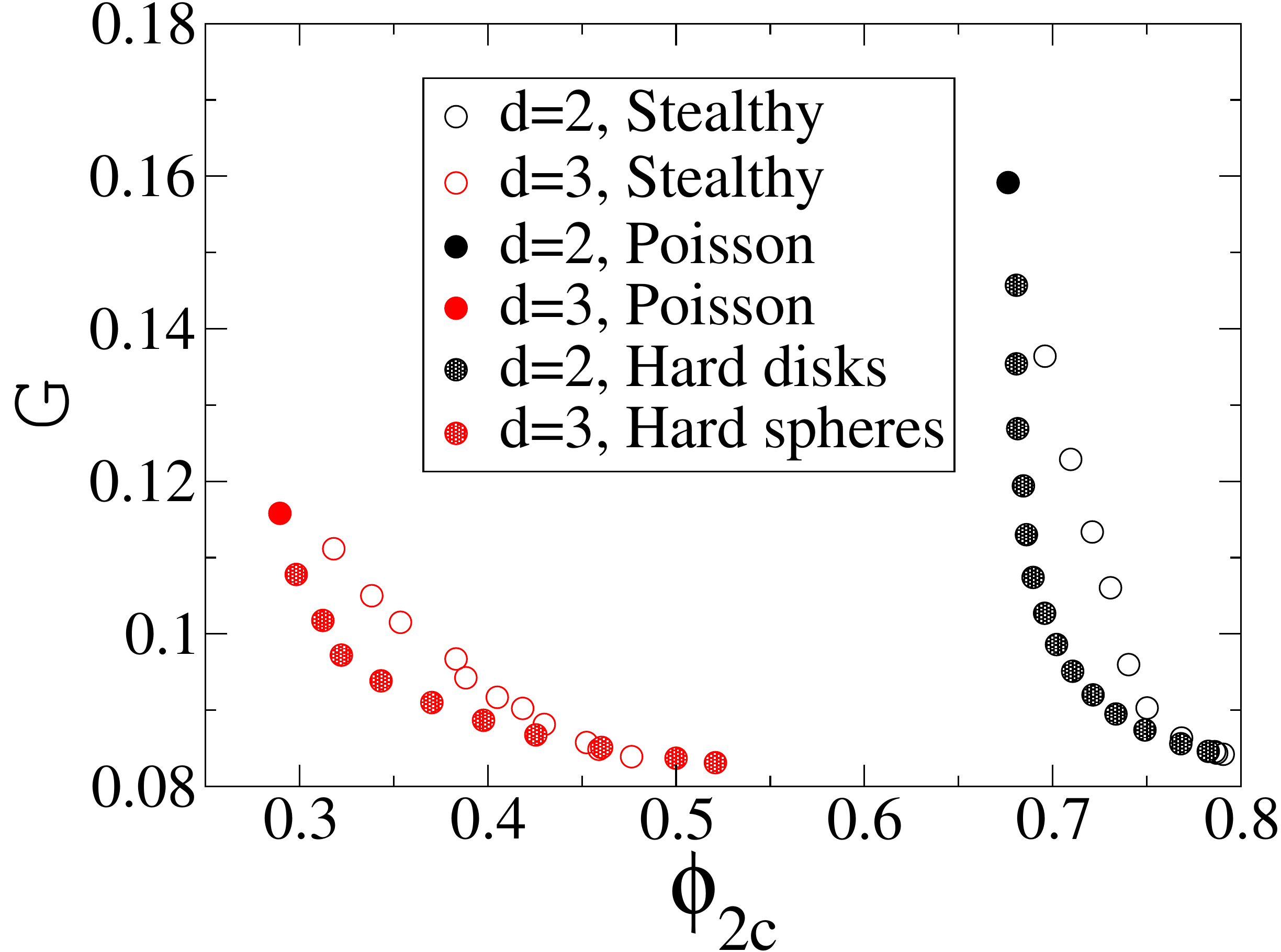}
\includegraphics[width=0.48\textwidth]{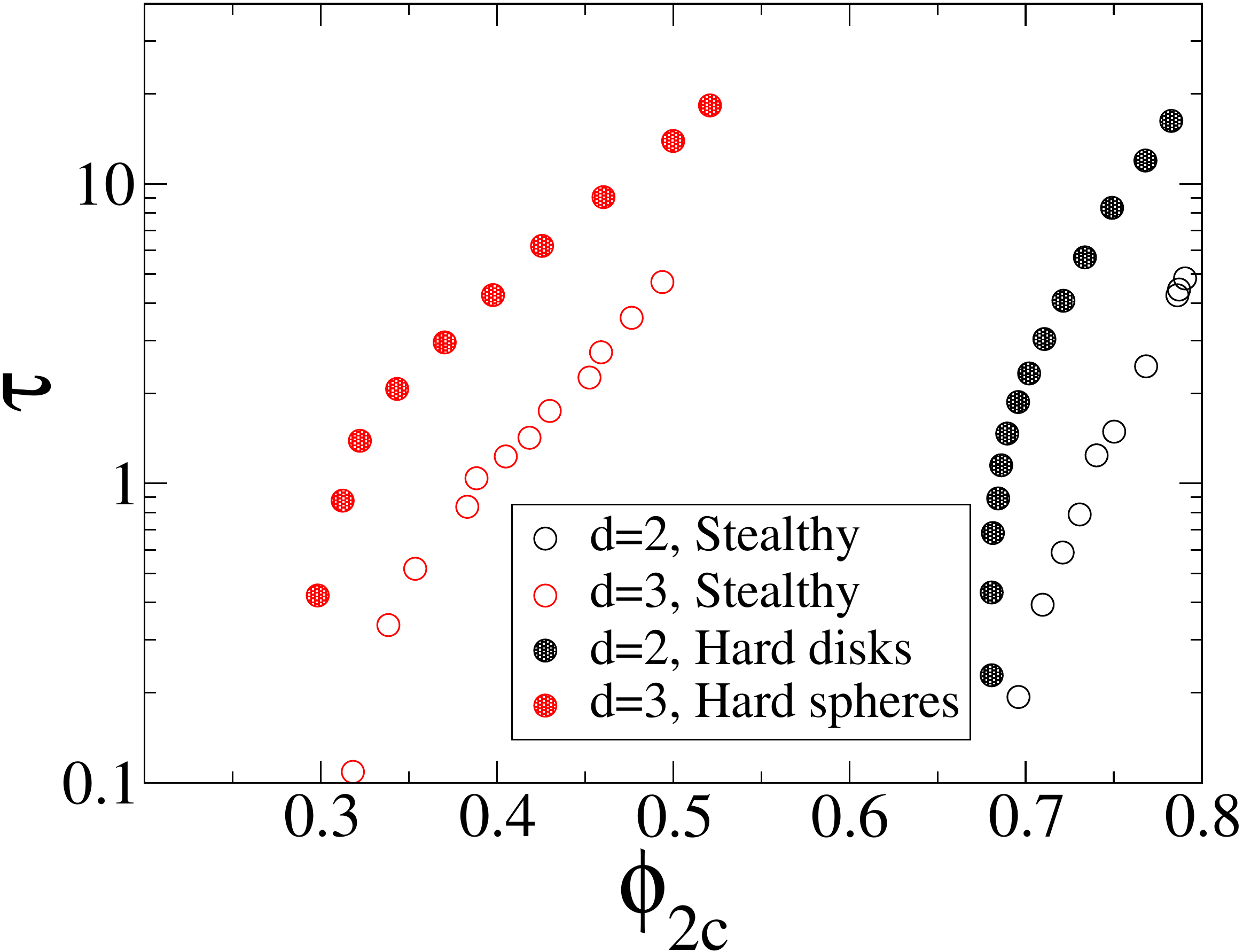}
\end{center}
\caption{Correlations between quantizer error $\mathcal{G}$, order metric $\tau$, and percolation volume fraction $\phi_c$ at unit number density.}
\label{correlations}
\end{figure}

In Fig.~\ref{correlations}, we explore the correlation between $\mathcal{G}$, $\tau$, and $\phi_c$, and compare that for entropically favored stealthy ground states with that for equilibrium disordered hard-sphere systems. It is interesting to note that these two systems behave very differently. At the same $\tau$, stealthy systems give lower values of $\mathcal G$, indicating $\mathcal G$ is more sensitive to long-range order than to short-range order. At the same $\tau$ or $\mathcal G$, stealthy systems give higher values of $\phi_{2c}$, indicating that $\phi_{2c}$ is even more sensitive to long-range order than to short-range order.

\section{Conclusions and discussion}
\label{conclusion}

In this work, we decorated stealthy disordered hyperuniform point configurations of different degrees of order with spheres of various radii, and computed several transport and structural properties of these decorated systems. The transport properties that we studied include effective diffusion coefficient $D_e$, mean survival time $T_{mean}$, survival probability $p(t)$, and principal relaxation time $T_1$. The structural properties examined include hyperuniformity and stealthiness, maximum packing fraction $\phi_p^{max}$, the void-exclusion probability $E_V$, the order metric $\tau$, and the percolation thresholds $\phi_{1c}$ and $\phi_{2c}$. 
We showed that the order metric 
$\tau$ is an exquisite detector of both short- and long-range translational order.
While all geometrical
and topological quantities are strongly correlated (positive correlation between $\phi_{2c}$ and $\tau$, and negative correlation between $\mathcal{G}$ and the former two quantities), the relation between the physical quantities are more complex: While $D_e$ increases as $\chi$ increases or as $\phi_2$ decreases, $T_{mean}$ and $T_1$ increases as $\chi$ decreases or as $\phi_2$ decreases. Therefore, there is no simple relationship between $D_e$ and $T_{mean}$ or $T_1$. Another reason why there is no such relationship is because if phase 1 ceases to percolate, then $D_e$ becomes zero but $T_{mean}$ and $T_1$ are still positive \cite{torquato2013random}.

Besides finding correlations between geometrical
and topological properties, we find that in the highly disordered ($\chi \ll 1$) regime, $T_1$ of our two-phase systems derived from decorated stealthy ground states is much lower than that of equilibrium hard-sphere system. Together with the void-exclusion probability, low $T_1$ suggests that the formation of large holes is strongly suppressed, even though the configuration appears completely disordered. 

In the higher order ($\chi \approx 0.5$) regime, 
$D_e$ of our disordered isotropic two-phase systems derived from decorated stealthy ground states is very close to the Hashin-Shtrikman upper bound for $0\le \phi_2<\phi_p^{max}$, where $\phi_p^{max}$ is the maximum packing fraction. Since such decorated systems maintain stealthiness if and only if $\phi_2<\phi_p^{max}$, our results suggest a connection between stealthiness and the ability to have a nearly optimal (maximal) $D_e$.
The fact that stealthy disordered two-phase media have nearly optimal $D_e$ could have practical implications, e.g., optimal and isotropic drug release from designed  nanoparticles. Although nearly optimal $D_e$ can also be achieved by lattice structures (i.e., periodic arrays of inclusions), the latter are always anisotropic. Thus, if one desires isotropic two-phase media with highest possible $D_e$ at a specific volume fraction, disordered stealthy two-phase media could be the best choice.

Disordered stealthy ground states are uncountably infinitely degenerate.\cite{zhang2015ground} The maximum packing fraction $\phi_p^{max}$ varies among configurations. In the future, it would be interesting to design algorithms that sample stealthy ground states with a bias toward configurations with higher $\phi_p^{max}$ values. With such an algorithm, one would be able to design isotropic two-phase media with nearly optimal $D_e$ with very high $\phi_2$.

Lastly, we would like to mention that although here we only study the diffusion problem of point Brownian particles, the diffusion problem of finite-sized Brownian particles has also been of interest.\cite{ashworth2015cell} It is noteworthy that our results can be trivially extended to the latter case. The diffusion of Brownian particles of radius $b$ among obstacles of radius $a$ is equivalent to the diffusion of point Brownian particles among obstacles of radius $a+b$. This mapping was previously exploited to quantify diffusion of finite-sized spheres in various models of porous media.\cite{kim1992diffusion}

Interestingly, one can relate the transport properties computed here ($D_e$, $T_{mean}$, and $T_1$) to different physical properties of the same systems via cross-property relations, including those that relate them to the elastic moduli, \cite{gibiansky1993link, gibiansky1996connection} as well as fluid permeability. \cite{avellaneda1991rigorous, torquato1990relationship} In future work, we will carry out such analyses.

\begin{acknowledgments}
The authors thank Duyu Chen and Jaeuk Kim for their careful reading of the manuscript.
\end{acknowledgments}

\appendix

\section{System sizes}
\label{number}
As discussed in Sec.~\ref{GeneratingStealthy}, it is nontrivial to choose the system size $N$ and parameter $\chi$, especially because one of our protocol to calculate the percolation threshold requires three different $N$'s for each $\chi$. We enumerated all possible choices of $N$'s and $\chi$'s for $N<1000$ and picked up some $\chi$ values that allow at least three different choices of $N$'s. Our choice of $N$ and $\chi$ in 2D and 3D are listed in Tables~\ref{N_2D} and \ref{N_3D}. It is desirable to consider values of $\chi$ higher than 0.4133... in 3D, but our enumeration did not find such a $\chi$ value that satisfies the above condition.
Therefore, we chose three more $N$'s that allow $\chi$ to be very close to $0.4598$ but makes $\chi$ differ in the fifth decimal place. See the caption for Table~\ref{N_3D} for details.
Except for the percolation threshold calculation, we only use the largest $N$ for each $\chi$ and $d$.


\begin{table}[h!]
\setlength{\tabcolsep}{12pt}
\caption{Our choice of parameter $\chi$'s, and the corresponding three different numbers of particles, $N_1$, $N_2$, and $N_3$ in 2D.}
\begin{tabular}{c c c c}
\hline
$\chi$ & $N_1$ & $N_2$ & $N_3$\\ \hline
0.05 & 151 & 451 & 751 \\
0.1 & 106 & 301 & 496 \\
0.15 & 101 & 311 & 501 \\
0.2 & 106 & 301 & 511 \\
0.3 & 101 & 301 & 511 \\
0.35 & 121 & 301 & 481 \\
0.4 & 106 & 271 & 511 \\
0.45 & 111 & 311 & 471 \\
0.465 & 101 & 301 & 501 \\
0.48 & 126 & 326 & 476 \\
\hline
\end{tabular}
\label{N_2D}
\end{table}

\begin{table}[h!]
\setlength{\tabcolsep}{12pt}
\caption{Our choice of parameter $\chi$'s, and the corresponding three different numbers of particles, $N_1$, $N_2$, and $N_3$  in 3D. The ``*'' mark indicates that $\chi$ values differ starting from fifth decimal place between the three choices of $N$.}
\begin{tabular}{c c c c}
\hline
$\chi$ & $N_1$ & $N_2$ & $N_3$\\ \hline
0.02 & 151 & 351 & 651 \\
0.0555... & 127 & 259 & 421 \\
0.0833... & 109 & 281 & 497 \\
0.1333... & 176 & 311 & 476 \\
0.1666... & 135 & 321 & 459 \\
0.2083... & 113 & 257 & 425 \\
0.2333... & 101 & 161 & 431 \\
0.2777... & 121 & 319 & 475 \\
0.3333... & 101 & 302 & 480 \\
0.3690... & 113 & 309 & 477 \\
0.4133... & 101 & 276 & 426 \\
0.4598...* & 167 & 383& 520\\
\hline
\end{tabular}
\label{N_3D}
\end{table}

\section{Properties of stealthy point configurations and decorated systems}
\label{properties}

In this section we tabulate all of the physical and geometrical properties of stealthy point configurations and decorated systems that we study in this paper (Tables III and IV).

\onecolumngrid
\begin{table}[H]
\setlength{\tabcolsep}{12pt}
\caption{Principal relaxation time $T_1$ at $\phi=0.2$ and $\phi=0.5$, order metric $\tau$, quantizer error $\mathcal G$, void-phase and particle-phase percolation volume fraction $\phi_{1c}$ and $\phi_{2c}$, and void-phase and particle-phase percolation radius $a_{1c}$ and $a_{2c}$ for different parameter $\chi$'s in 2D.}
\begin{tabular}{c c c c c c c}
\hline
$\chi$  &   $T_1$ ($\phi=0.2$)    &    $T_1$, ($\phi=0.5$)    &    $\tau$   &  $\mathcal{ G }$ &     $\phi_{1c}=1-\phi_{2c}$ &  $a_{1c}=a_{2c}$\\

0.05 &   0.2753&    0.1858&    0.193 &  0.1364 & 0.3041& 0.5840\\
0.1 &    0.1964&    0.1214&     0.393 &  0.1229 &     0.2904& 0.5692\\
0.15 &   0.1554&    0.0886&    0.588 & 0.1134 &    0.2790&  0.5564\\
0.2 &    0.1291&     0.0704&    0.787 & 0.106 &      0.2693& 0.5448\\
0.3 &    0.1015&    0.0502&    1.241 &0.096 &      0.2597& 0.5238\\
0.35 &   0.0842&    0.0408&    1.488 &0.0903 &     0.2498& 0.5117\\
0.4 &    0.0710&     0.0331&    2.459 &0.0863 &     0.2316& 0.5071\\
0.45 &   0.0657&    0.0282&    4.248 &0.0845 &     0.2139& 0.5083\\
0.465 &  0.0645&    0.0274&    4.443 &0.0844 &     0.2129& 0.5082\\
0.48 &   0.0635&    0.0272&     4.839 &       0.0842 &     0.2095 & 0.5090\\

\hline
\end{tabular}
\label{properties_2D}
\end{table}

\begin{table}[H]
\setlength{\tabcolsep}{12pt}
\caption{Same as above, except for 3D.}
\begin{tabular}{c c c c c c c c c}
\hline
$\chi$  &   $T_1$ ($\phi=0.2$)    &   $T_1$ ($\phi=0.5$)       &    $\tau$   &  $\mathcal{ G }$ &     $\phi_{2c}$ & $\phi_{1c}$ &  $a_{2c}$ & $a_{1c}$\\
0.02 &   0.1572&     0.0938&        0.107&     0.1111&     0.3182 &       0.0299 &0.4483&0.8970\\
0.0555 & 0.1221 &   0.0647&        0.336 &0.1050& 0.3384  &0.0261  &      0.4536&0.8572\\
0.0833 & 0.1061 &   0.0551&        0.518&0.1015&0.3536  &0.0242  &      0.4583&0.8351\\
0.1333 & 0.0911 &   0.0433&        0.835&0.0967&0.3832  &0.0225  &      0.4672&0.8019\\
0.1666 & 0.0826 &   0.0387&        1.040&0.0942&0.3884  &0.0205& 0.4664&0.7878\\
0.2083 & 0.0779 &   0.0342&        1.231&0.0917&0.405  & 0.0196  &      0.4702&0.7700\\
0.2333 & 0.0722 &   0.0317&        1.420 &      0.0902&0.4185  &0.0181  &      0.4739&0.7623\\
0.2777 & 0.0668 &   0.0289&        1.745 &      0.0881&0.4300  &  0.0192& 0.4753&0.7446\\
0.3333 & 0.0612 &   0.0254&        2.258 &      0.0858&0.4525  &0.0199  &      0.4804&0.7253\\
0.369 &  0.0593 &   0.0236&        2.740 &      0.0849&0.4591  &0.0198  &      0.4814&0.7166\\
0.4133 & 0.0561 &   0.0213&        3.575 &      0.0839&0.4764  &0.0194& 0.4864&0.7082\\
0.4598 & 0.0537 &   0.0202&        4.704 &      0.0823&0.4939  &0.0201& 0.4917&0.6992\\
\hline
\end{tabular}
\label{properties_3D}
\end{table}

\twocolumngrid

\end{document}